\documentclass[referee,sn-nature]{sn-jnl}

\usepackage{graphicx}
\usepackage{multirow}
\usepackage{amsmath,amssymb,amsfonts}
\usepackage{amsthm}
\usepackage{mathrsfs}
\usepackage[title]{appendix}
\usepackage{xcolor}
\usepackage{textcomp}
\usepackage{manyfoot}
\usepackage{booktabs}
\usepackage{algorithm}
\usepackage{algorithmicx}
\usepackage{algpseudocode}
\usepackage{listings}
\usepackage{xcolor}
\usepackage{lmodern}
\usepackage{verbatim}

\begin{document}


\title[Article Title]{Unlocking Quantum Control and Multi-Order Correlations via Terahertz Two-Dimensional Coherent Spectroscopy}


\author[1,2]{\fnm{Chuankun} \sur{Huang}}

\author[1]{\fnm{Martin} \sur{Mootz}}

\author[1]{\fnm{Liang} \sur{Luo}}

\author[3]{\fnm{Ilias E.} \sur{Perakis}}

\author*[1,2]{\fnm{Jigang} \sur{Wang}}
\email{jgwang@ameslab.gov, jgwang@iastate.edu}

\affil[1]{\orgdiv{Ames National Laboratory}, \orgname{U.S. Department of Energy}, \\ \orgaddress{\city{Ames}, \postcode{50011}, \state{IA}, \country{USA}}}

\affil[2]{\orgdiv{Department of Physics and Astronomy}, \orgname{Iowa State University}, \\ \orgaddress{\city{Ames}, \postcode{50011}, \state{IA}, \country{USA}}}

\affil[3]{\orgdiv{Department of Physics}, \orgname{University of Alabama at Birmingham}, \\ \orgaddress{ \city{Birmingham}, \postcode{35294-1170}, \state{AL}, \country{USA}}}

\abstract{Terahertz two-dimensional coherent spectroscopy (THz-2DCS) is transforming our ability to probe, visualize, and control quantum materials far from equilibrium. This emerging technique brings multi-dimensional resolution to the ultrafast dynamics of nonequilibrium phases of matter, enabling new capabilities demanding precise coherent control and measurement of many-body dynamics and multi-order correlations.  By mapping complex  excitations across time and frequency dimensions, THz-2DCS delivers coherence tomography of driven quantum matter, thus revealing hidden excitation pathways, measuring higher-order nonlinear response functions, disentangling various quantum pathways, capturing collective modes on ultrafast timescales and at terahertz frequencies. These experimental features  frequently remain obscured in traditional single-particle measurements, ultrafast spectroscopy techniques, and equilibrium-based probes.  This Technical Review traces the early development of THz-2DCS and showcases significant recent progress in leveraging this technique to probe and manipulate quantum material properties, including nonequilibrium superconductivity, nonlinear magnonics, dynamical topological phases, and the detection of novel excitations and exotic collective modes with potential technological impact. Looking forward, we identify critical opportunities in advancing THz-2DCS instrumentation and experimental strategies that are shaping future applications in THz optoelectronics, quantum information processing, and sensing.}



\maketitle

\pagebreak

\begingroup
\let\clearpage\relax 

\section*{Introduction}

Over the past few decades, transformative advances in condensed matter and materials physics (CMMP) have accelerated our understanding and applications of quantum coherence~\cite{Basov2017,Awschalom2018}. Advancing the convergence of Quantum Information Science (QIS) and CMMP hinges on quantum functionalities that require studying quantum dynamics and achieving coherent control in materials~\cite{Georgescu2014,Preskill2018}. While significant progress has been made across current QIS platforms---including cold-atom quantum simulators~\cite{Gross2017}, trapped-ion systems~\cite{Foss-Feig2025}, and superconducting qubit processors driven by coherent electromagnetic waves~\cite{Warner2025}---terahertz (THz) nonlinear spectroscopy~\cite{Elsaesser2Dbook,Nicoletti:16}, particularly two-dimensional coherent spectroscopy (2DCS), is now emerging as a complementary approach for achieving coherent control~\cite{bristow2009versatile,Cundiff2013,li2006many,yang2007two,li2013unraveling,cowan2005ultrafast,dahms2017large,Yang2023} and probing superconducting quantum dynamics~\cite{yang2018,yang2019lightwave,Luo2023} at higher frequencies and elevated temperatures.

By dynamically manipulating many-body correlations and collective modes in quantum materials at higher THz frequencies, THz-2DCS expands the frontier of coherent control capabilities,  surpassing the GHz low frequency limitations of conventional superconducting quantum circuits and other QIS platforms~\cite{Preskill2018}. In this review, we highlight the pivotal role of  THz-2DCS in probing nonlinear quantum dynamics, and in enabling  coherent control of complex quantum materials. By introducing a second scanning dimension beyond conventional frequency and time-domain measurements--linked to the relative phase of the electromagnetic fields--THz-2DCS 
enables correlation tomography with spectral-temporal resolution, disentangling multiple quantum pathways contributing at the same frequency. We highlight two key advantages.

\begin{figure}[!ht] 
\centering 
\includegraphics[width=\textwidth]{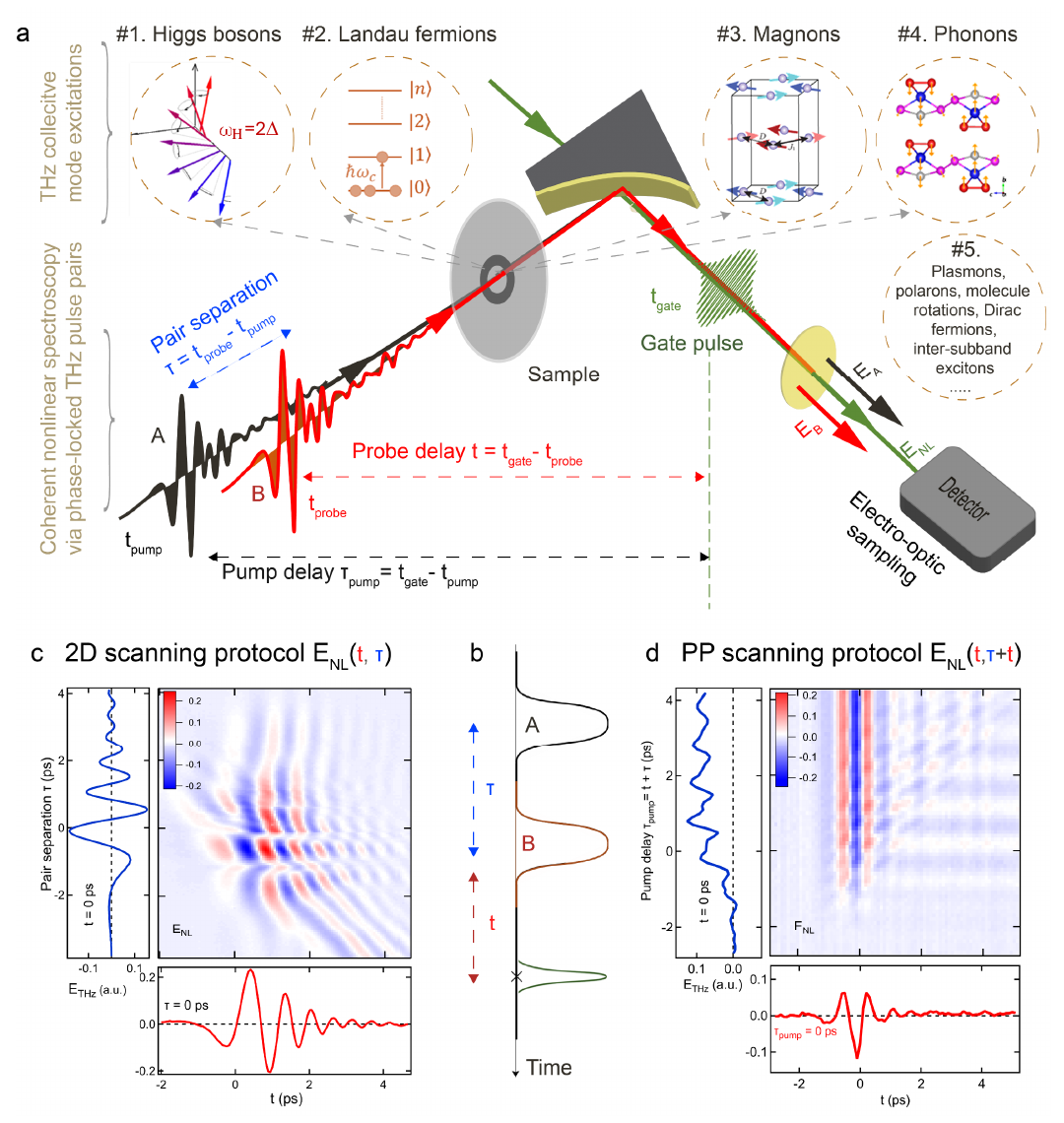} 
\caption{\textbf{Technical development and evolution of THz-2DCS.} 
\textbf{a}, Schematic of THz nonlinear spectroscopy, particularly two-dimensional coherent spectroscopy, based on two phase-locked THz pulses, A (black) and B (red), as a function of gate time $t_\text{gate}$ and inter-pulse delay $\tau$, with  nonlinear emission detected via phase-resolve electro-optic sampling. This scheme showcases the unique capabilities of THz-2DCS for probing quantum coherence and correlations of THz collective modes (indicated by dashed circles labeled \#1--\#5) in quantum materials.
\textbf{b}, Time sequences of THz pulses A and B relative to the gate pulse. 
\textbf{c}, Representative THz nonlinear emission data obtained using a two-dimensional (2D) scanning protocol in an FeAs superconductor~\cite{Luo2023}.
\textbf{d}, THz nonlinear emission data acquired via a pump–probe (PP) scanning protocol in an Nb$_3$Sn superconductor~\cite{yang2019lightwave}.
} 
\label{fig_intro}
\end{figure}

First, THz-2DCS captures the correlation dynamics of two excitations, rather than probing individual excitations. This is critical for disentangling and controlling the complex properties of quantum materials. Beyond simply probing the properties of equilibrium  states, THz-2DCS enables more detailed excitation-detection measurements of coherence, symmetry, correlations, nonlinear couplings, and non-perturbative responses under far-from-equilibrium conditions. 
Therefore, the pulse-pair excitations with finely controlled amplitudes and phases enable both the realization and deeper understanding of the desired final states.
These enhanced capabilities open new avenues for overcoming tuning limitations in materials design, and for engineering highly coherent, nonlinear quantum systems, as required for applications in both fundamental quantum materials research and QIS. Such dynamical control is often inaccessible with conventional characterization and control methods, which rely on slow thermodynamic tuning or linear responses.

Second, THz-2DCS enables the disentanglement and control of emergent quantum phases characterized by exotic collective modes and multi-order correlations. This capability is partially achieved by nonlinearly modulating the ``soft" properties of the equilibrium state, and by simultaneously measuring both amplitude and phase dynamics of the emitted light fields. Therefore, THz-2DCS reshapes our understanding of coherences and correlations in quantum materials, both resolving and selectively controlling intertwined collective modes and separating excitation pathways in a two-dimensional plane. Such responses are often obscured in traditional one-dimensional (1D) pump-probe techniques and single-particle spectroscopies, where multiple correlated excitations overlap and contribute indistinguishably to the same spectral-temporal signals.


Over the past decade, experimental and theoretical THz-2DCS studies have yielded significant discoveries for quantum material research. As illustrated in Fig.~\ref{fig_intro}\textbf{a}, collective modes, such as Higgs bosons, Landau fermions, magnons, and phonons (dashed circles labeled \#1--\#4), are observed as prominent spectral peaks at distinct 2D frequencies, clearly separated from other excitations. Such features, along with recent advances involving additional collective modes and elementary excitations (labeled \#5), lay the foundation for extending THz-2DCS from early fundamental studies of semiconductors to achieve coherent control over superconducting, magnetic, topological, and other emergent quantum phases. 
The ultimate goal of these studies is to overcome the materials bottleneck that limit the development of highly coherent quantum computing and sensing technologies.


The general framework of THz nonlinear spectroscopy of quantum materials using two phase-locked THz pulses (black and red) in a phase-resolved detection configuration and with a third sampling pulse (green) is illustrated in Fig~\ref{fig_intro}\textbf{a}. The two applied THz laser pulses, separated by inter-pulse delay time $\tau$, are denoted as $E_\text{A}$ (black) and $E_\text{B}$ (red). The material's nonlinear response to these pulses generates the nonlinear emission signal $E_\text{NL}=E_\text{A+B}-E_\text{A}-E_\text{B}$, where $E_\text{A+B}$ is the response when both pulses are applied together, and $E_\text{A}$ and $E_\text{B}$ are the responses to each pulse individually. This coherent THz emission signal can be detected via electro-optic sampling (EOS), which enables the simultaneous measurement of both the amplitude and the phase of the emitted electric field. The observables extracted from  this nonlinear signal are highly versatile and material-specific,  encompassing  polarizations, currents, magnetizations, collective modes, and  elementary excitations of quantum materials.

There are two complementary scanning protocols for THz coherent nonlinear spectroscopy, shown in Figs.~\ref{fig_intro}\textbf{b}--\textbf{d}. These protocols use three pulses (Fig.~\ref{fig_intro}\textbf{a}): pump pulse A at time $t_\text{pump}$, probe pulse B at time $t_{\text{probe}}$, and optical gating pulse at time $t_{\text{gate}}$. The measured signals depend on three controllable time delays illustrated in Figs.~\ref{fig_intro}\textbf{a} and \ref{fig_intro}\textbf{b}: time separation of pulses A and B,  $\tau = t_\mathrm{probe}-t_\mathrm{pump}$, coherent emission time  $t=t_\mathrm{gate}-t_\mathrm{probe}$, and pump-induced evolution time $\tau_\mathrm{pump}=t+\tau$ (or $t_\mathrm{gate}-t_\mathrm{pump}$). 
Representative THz-2DCS raw data for the two scanning protocols are shown in Fig.~\ref{fig_intro}\textbf{c} (2D scanning protocol) and Fig.~\ref{fig_intro}\textbf{d} (pump-probe (PP) scanning protocol). These two protocols can be distinguished by  the relative timings of the THz and gate pulses (Fig.~\ref{fig_intro}\textbf{b}), which are controlled by the above time delays. 
Specifically, the 2D scanning protocol (Fig.~\ref{fig_intro}\textbf{c}) detects the coherent nonlinear THz emission at fixed pulse-pair separation $\tau$,  measuring the two-time nonlinear response function $E_\text{NL}(t, \tau)$.
The PP scanning protocol (Fig.~\ref{fig_intro}\textbf{d}) detects two-pulse interference at fixed time $t$, or excited-state dynamics at fixed pump delay $\tau_\mathrm{pump}=t+\tau$. The nonlinear signal $E_\text{NL}(t, \tau_\mathrm{pump})$ that mixes $t$ and $\tau$ is measured. The PP method yields the same response function $E_\text{NL}(t, \tau)$ as the full 2D scanning protocol, up to an appropriate transformation of the time axes, and vice versa. 
Both scanning methods are complementary and can be used synergistically to visualize and characterize material nonlinear features. For example, the 2D spectra directly measure collective modes and enable coherent control, as they can capture the full nonlinear emission $E_\text{NL}(t, \tau)$ at each fixed $\tau$, while the PP method provides a direct means to study the material dynamical response functions, such as ultrafast conductivity $\sigma_\text{NL}(t, \tau_{\text{pump}})$ and permittivity $\epsilon_\text{NL}(t, \tau_{\text{pump}})$. This method avoids artifacts like ``perturbed free induction decay", allowing for cleaner extraction of the material's optical response functions convoluted in the 2D signals. 

Boxes~1 and 2 illustrate how THz-2DCS uniquely probes collective excitation dynamics and separates multi-order quantum pathways via the nonlinear signal $E_\text{NL}(t, \tau)$ and the corresponding 2D spectra. When applied to quantum materials, THz-2DCS offers additional strategic advantages which demand accurate interpretation and theoretical modeling:
\begin{itemize}
    \item \textit{\textbf{Coherence tomography, multi-order correlation, and quantum pathway disentanglement:}} THz-2DCS simultaneously accesses order parameters, collective modes, and their associated coherences and fluctuations through the two-dimensional frequency space. This capability enables the resolution of competing quantum pathways and the disentanglement of multi-order correlations~\cite{Luo2023,Lu2019,Liu2025,Huang2023}.


   \item \textit{\textbf{Control of supercurrents and other anomalous quantum transport:}} Excitation of superconductors by phase-locked THz pulse pairs enables the direct manipulation of persistent supercurrents by generating and controlling superfluid momentum $p_\text{S}(t)$. This process allows control of Cooper pair center-of-mass momentum over timescales exceeding the pulse duration, offering a powerful mechanism for coherent quantum state control and light-induced symmetry breaking~\cite{yang2019lightwave, vaswani2019discovery,Mootz2020}. The same approach can be extended to dissipationless currents in topological states and other quantum systems \cite{luo,vasw2020,Yang2020}.

    
    \item \textit{\textbf{Coherently enhanced sensing efficiency and fidelity:}} THz-2DCS enables the separation of distinct coherent nonlinear signals from detrimental artifacts by identifying them at background-free positions within the 2D frequency plane. The coherent nonlinear emission induced by phase-locked laser pulse-pairs exhibits well-defined spectral peaks that predominantly arise from coherent oscillations of collective modes~\cite{Huang2024} and from non-perturbative nonlinear responses~\cite{Luo2023,mootz2023multidimensional}. These peaks are well-separated from conventional peaks, e.g., dominated by quasiparticle excitations following incident pulse frequency. 
\end{itemize}



THz-2DCS studies have advanced rapidly in recent years, expanding beyond semiconductors to encompass a broad range of quantum materials, including superconductors, magnetic materials, topological systems, and other strongly correlated electronic materials. Table~\ref{table} summarizes recent  applications of THz-2DCS  across these emerging platforms, highlighting the technique's utility in addressing key research challenges--ranging from transient high-temperature superconductivity and nonequilibrium topological phases to nonlinear magnonics and phononics. 
These studies employ THz pulses with field strengths up to 1--10~MV/cm and photon energies in the range of $\sim$ 0.1--100 meV (second column),  tailored to probe materials' quantum states  and characterize them by measuring their collective modes (third column). The experimental protocols (fourth column) vary by using  both single-cycle (broadband) and multi-cycle (narrowband, tunable) THz pulses, as well as by using  resonant or off-resonant excitation relative to energy gaps determined by correlations in quantum materials, among others. The experimental protocols  for achieving  suitable spectral-temporal resolution are implemented to probe collective modes and address outstanding questions across different fields (fifth column), as listed in Table~\ref{table}. 

\begin{figure}[!ht]
\renewcommand{\figurename}{Table}
\setcounter{figure}{0}
\centering 
\includegraphics[width=\textwidth]{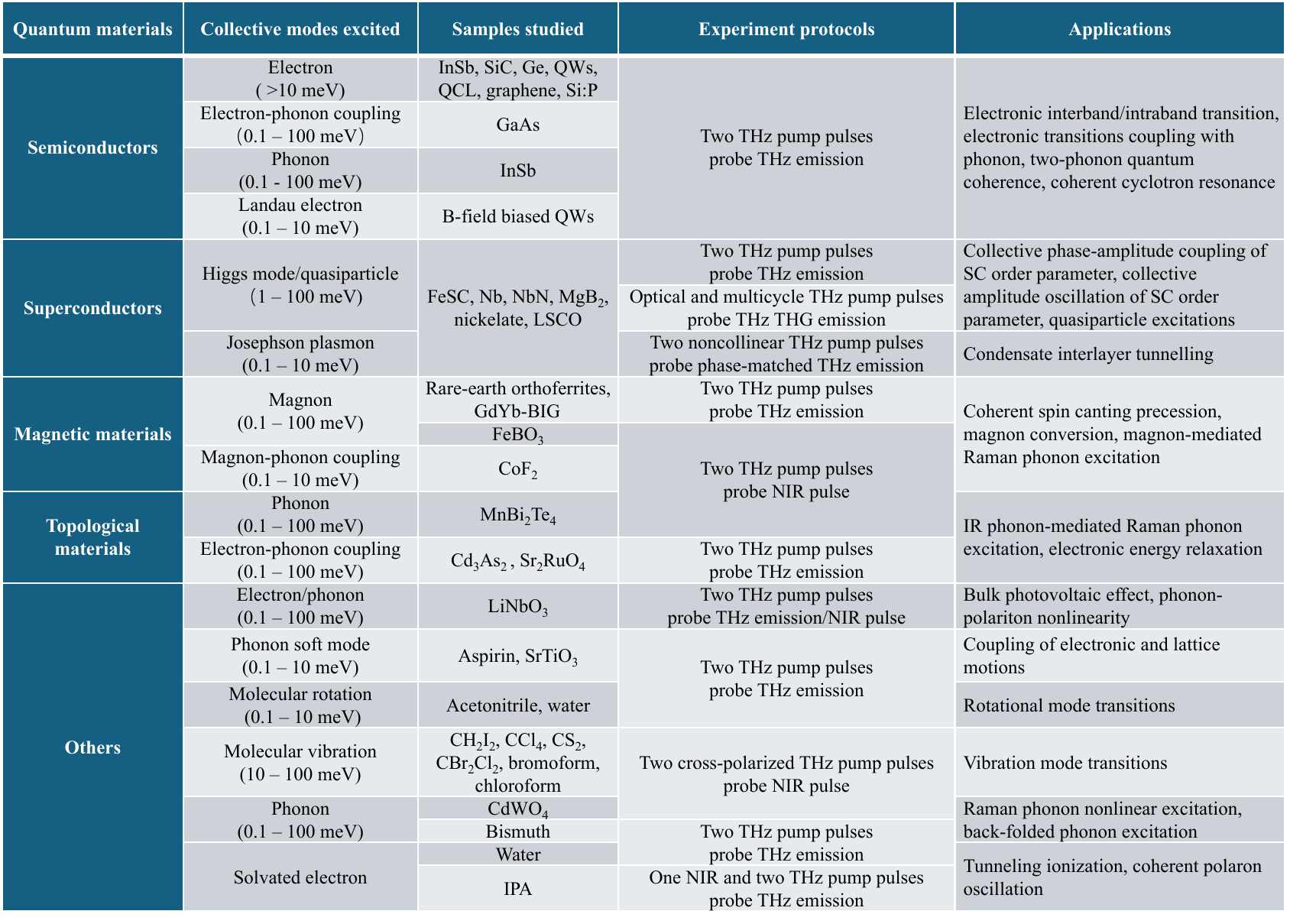} 
\caption{An overview of THz-2DCS experiments in various quantum materials, including semiconductors~\cite{Kuehn2009,Kuehn2011QW1,Junginger2012,Woerner2013,Woerner2019,Ghalgaoui2020GaAs,Woerner2021,Kuehn2011QW2,Bowlan2014,Somma2016InSb,Somma2016beyond,Houver2019,Markmann2020,Raab2019,Raab2020,Tarekegne2020,Mahmood2021,Riepl2021,Gill2024,Elsaesser2015,Maag2016,Mornhinweg2021}, magnetic materials~\cite{Pashkin2013,Lu2017,Huang2024,Dutta2025,Mashkovich2021,Grishunin2023,Blank2023Spin,Zhang2024Coup,Zhang2024Down,Zhang2024Up,Leenders2024,Zhang2024Spin},
superconductors~\cite{Luo2023,Huang2023, Katsumi2024MgB2,Kim2024,Katsumi2024NbN,Liu2024,Salvador2024,chaudhuri2025,Cheng2025}, topological materials~\cite{Blank2023Phonon,Bhandia2024,Barbalas2025}, ferroelectric materials~\cite{Somma2014,Pal2021,Lin2022}, molecular systems~\cite{Reimann2021,Folpini2017,Lu2016,Zhang2021,Allodi2015,Finneran2016,Finneran2017,Magdau2019,Mead2020,Ghalgaoui2020Water,Runge2023Sol}, and others~\cite{Johnson2019,Biggs2023,Runge2023Bis}. (N)IR: (near-)infrared; THG: third-harmonic generation.} 
\label{table}
\renewcommand{\figurename}{Fig.}
\setcounter{figure}{1}
\end{figure}

Given the rapid expansion of THz-2DCS applications across a broad range of correlated quantum and topological materials, this comprehensive Technical Review synthesizes the field’s current state, identifies emerging opportunities, and charts future directions. 
The organization of this review follows the structure outlined in Table~\ref{table}, summarizing the evolving role of THz-2DCS in various materials research. 
We conclude with an outlook that links THz-2DCS to open questions in condensed matter and quantum materials physics, the development of novel coherent nonlinear THz spectroscopy under extreme conditions, and emerging directions in THz nonlinear spectroscopy at nanoscale. Altogether, this Technical Review aims to serve as both a reference and a roadmap for leveraging THz-2DCS to engineer exotic quantum coherence and entanglement, advancing future technologies for quantum control, computation, and sensing.


\pagebreak
\setcounter{figure}{0}
\subsection*{Box 1 Terahertz two-dimensional coherent spectroscopy}

\begin{figure}[!ht]
\centering 
\includegraphics[width=\textwidth]{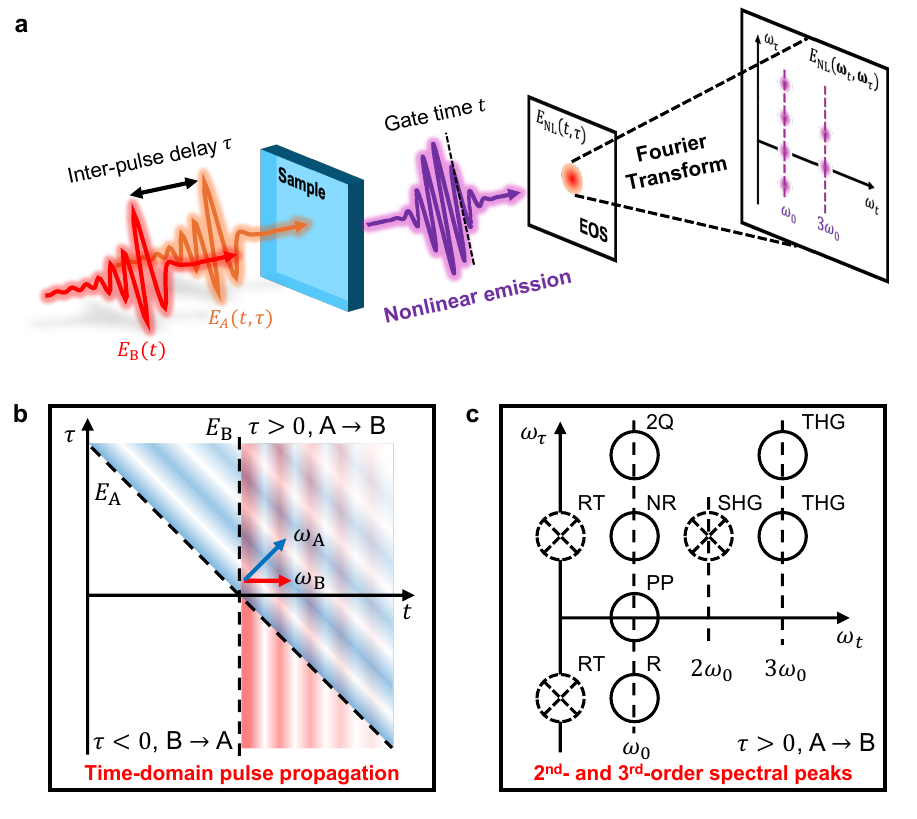} 
\label{box1}
\end{figure}

THz-2DCS is a powerful tool for investigating nonlinear light-matter interactions in quantum materials~\cite{Lu2019,Liu2025}. Utilizing phase-locked THz pulse sequences, it enables access to collective excitation dynamics, revealing  quantum correlations and unconventional quasiparticle interactions hidden in conventional spectroscopies.

In a typical THz-2DCS experiment, two collinear THz pulses, $E_\text{A}$ and $E_\text{B}$, separated by an inter-pulse delay $\tau$, excite the sample (panel~\textbf{a}). The resulting nonlinear emission is detected via EOS, providing phase-resolved nonlinear signals. Panel~\textbf{b} shows the time-domain representation of the two pulses as functions of gate time $t$ and inter-pulse delay $\tau$. Blue and red arrows indicate the wavevectors $\omega_\text{A}$ and $\omega_\text{B}$, while shaded regions represent their temporal wavefronts. The nonlinear response depends on pulse order: for $\tau > 0$, pulse A precedes B (top half of panel~\textbf{b}); for $\tau < 0$, the order is reversed (bottom half). The induced nonlinear signal arises either during the temporal overlap of the pulses or after their interaction, as a result of subsequent coherent evolution, thereby encoding information about quantum coherence and interactions within the system.

In the perturbative regime, nonlinear processes contributing to the 2D spectra are identified by expanding $E_\text{NL}(t,\tau)$ in powers of the applied electric fields~\cite{mukamel1995principles,Wan2019,Nandkishore2021}. Assuming impulsive pulses, the electric field is
\begin{align}
	E(t) = E_1 \,\delta(t) + E_2\,\delta(t-\tau)\,,
\end{align}
where $E_1$ and $E_2$ are the effective pulse areas centered at $t = 0$ and $t = \tau$, respectively. The nonlinear field transmitted along direction $\alpha$, measured at time $t+\tau$, can be expressed through $n$th-order nonlinear susceptibilities $\chi^{(n)}$~\cite{Wan2019,Choi2020,Nandkishore2021}:
\begin{align}
	&E_\mathrm{NL}^\alpha(t+\tau) 
	\equiv E^\alpha_\mathrm{AB}(t+\tau)-E^\alpha_\mathrm{A}(t+\tau)-E^\alpha_\mathrm{B}(t+\tau) \nonumber \\
	&= \chi^{(2)}_{\alpha\beta\gamma}(t,\tau)\,S^\beta_\text{A} S^\gamma_\text{B} +\chi^{(3)}_{\alpha\beta\gamma\delta}(t,\tau,0)\,(S^\beta_\text{A})^2 S^\gamma_\text{B} +\chi^{(3)}_{\alpha\beta\gamma\delta}(t,0,\tau)\,S^\beta_\text{A} (S^\gamma_\text{B})^2 +\mathcal{O}(E^4)\,.
 \label{eq:Mexp}
\end{align}
The second-order susceptibility is expressed through time-ordered quantum correlation functions:
\begin{align}
	\chi^{(2)}_{\alpha\beta\gamma}(t,\tau) = \frac{i^2}{\hbar^2} \theta(t)\theta(\tau) \langle [[\hat{O}^\alpha(t+\tau), \hat{O}^\beta(\tau)], \hat{O}^\gamma(0)] \rangle\,,
\end{align}
where $\hat{O}^\mu$ denotes a generalized observable associated with the emitted signal, such as polarization, current, or magnetization. The third-order susceptibilities read
\begin{align}
\label{eq:chi3}
	\chi^{(3)}_{\alpha\beta\gamma\delta}(t,\tau,0) &= \frac{(-i)^3}{\hbar^3} \theta(t)\theta(\tau)\theta(0)\langle [[[ \hat{O}^\alpha(t+\tau), \hat{O}^\beta(\tau)], \hat{O}^\gamma(0)], \hat{O}^\delta(-\tau)] \rangle\,, \\
	\chi^{(3)}_{\alpha\beta\gamma\delta}(t,0,\tau) &= \frac{(-i)^3}{\hbar^3} \theta(t)\theta(0)\theta(\tau)\langle [[[ \hat{O}^\alpha(t+\tau), \hat{O}^\beta(0)], \hat{O}^\gamma(\tau)], \hat{O}^\delta(\tau)] \rangle\,.
\end{align}
These susceptibilities characterize the material’s intrinsic nonlinear response, reflecting its many-body interactions, quantum coherence, and symmetry constraints. To analyze the THz-2DCS spectra, the time-domain susceptibilities are transformed into the frequency domain via a two-dimensional Fourier transform (2DFT) with respect to $\tau$ and $t$, yielding spectra as functions of excitation ($\omega_\tau$) and detection ($\omega_t$) frequencies. The resulting features--located at specific $(\omega_t, \omega_\tau)$ coordinates--trace resonant pathways dictated by $\chi^{(n)}$, including second-order effects like rectification (RT) and second-harmonic generation (SHG), and third-order processes such as third-harmonic generation (THG), rephasing (R), non-rephasing (NR), pump-probe (PP), and two-quantum coherences (2Q). Thus, THz-2DCS directly reveals the underlying susceptibilities and, therefore, the dynamics and correlations of quantum materials.

The measured 2D spectra provide rich information about nonlinear interactions, quantum coherence and the underlying structure of the quantum material state. Panel~\textbf{c} shows a representative THz-2DCS spectrum under perturbative excitation conditions, highlighting nonlinearities up to third order. It is obtained by applying a 2DFT to the $\tau > 0$ region (top half of panel~\textbf{b}). PP, NR, R, and 2Q signals emerge at distinct $\omega_\tau$ positions at fixed $\omega_t = \omega_0$ and THG at $\omega_t = 3\omega_0$, demonstrating a full set of third-order nonlinear responses (solid circles in panel~\textbf{c}). In the perturbative regime, $\omega_0$ typically corresponds to the laser central frequency;  more generally, it can represent a laser-induced mode frequency in systems with ``soft" excitations and collective modes modified by coherent THz excitation. 
Second-order nonlinear signals (dashed circles in panel~\textbf{c}), such as RT and SHG, may also appear, serving as signatures of symmetry breaking in materials lacking inversion symmetry.

\pagebreak
\subsection*{Box 2 THz-2DCS: A unique window into coherence and dynamics of quantum materials}

\begin{figure}[!ht]
\centering 
\includegraphics[width=\textwidth]{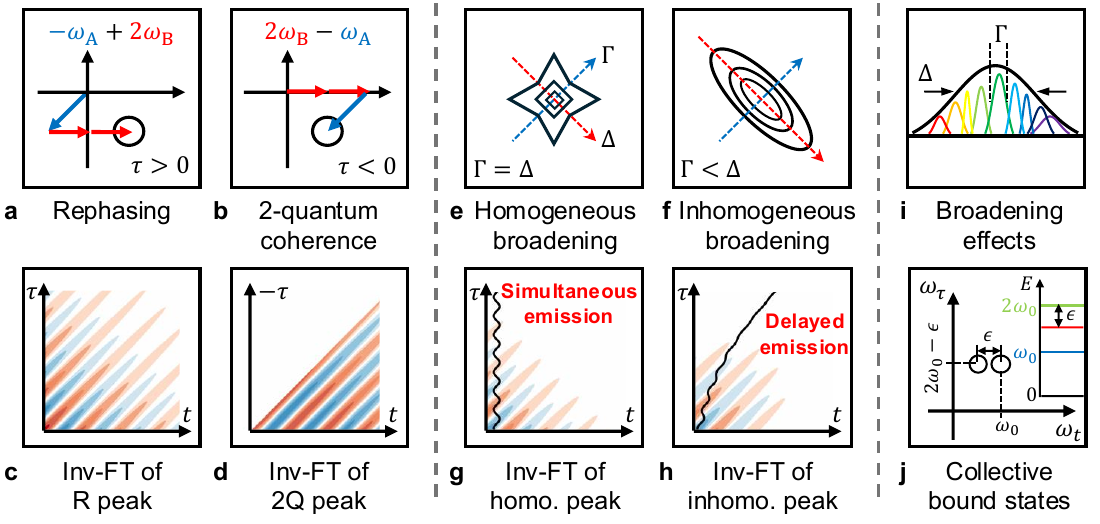} 
\label{box2}
\end{figure}

A key strength of THz-2DCS is its ability to disentangle spectral contributions arising from distinct interaction pathways. Notably, peaks at the same frequencies in the 2D spectrum can originate from different pulse interaction sequences.
For example, a rephasing signal ($\tau > 0$, panel~\textbf{a}) corresponds to an interaction sequence where pulse A excites the system first, followed by two interactions with pulse B, while a two-quantum coherence signal ($\tau < 0$, panel~\textbf{b}) arises when two interactions from pulse B precede a single interaction from pulse A. Applying inverse Fourier transforms (inv-FT) to selected regions of the 2D frequency spectrum allows isolation of these signals in the time domain (panels~\textbf{c} and \textbf{d}), enabling further separation of nonlinear processes and direct access to coherence lifetimes.

Beyond spectral resolution, THz-2DCS provides insights into distinguishing homogeneous (panel~\textbf{e}) from inhomogeneous (panel~\textbf{f}) broadening mechanisms. As illustrated in panel~\textbf{i}, a broad signal of spectral width $\Delta$ (black arrows) can comprise multiple narrow-band transitions with linewidth $\Gamma$ (dashed lines). For homogeneous broadening, associated with intrinsic decoherence, spectral features are elongated equally along both diagonal (red arrow, width $\Delta$) and off-diagonal (blue arrow, width $\Gamma$) directions, such that $\Gamma=\Delta$ (panel~\textbf{e}). In contrast, inhomogeneous broadening from disorder results in elliptical spectral shapes with $\Gamma < \Delta$ (panel~\textbf{f}). Inverse Fourier transforms of the frequency-domain signals further distinguish these two effects: in homogeneous systems (panel~\textbf{g}), the signal is emitted immediately following the second pulse (free-induction decay, black wiggled line), while in inhomogeneous systems (panel~\textbf{h}), delayed emission (black wiggled line) arises due to dephasing.

Beyond probing an ensemble of uncorrelated two-level systems displaying homogeneous and inhomogeneous broadening,  a particularly compelling application of THz-2DCS lies in its sensitivity to quantum material anharmonicities from fundamental interactions, exemplified by the emergence of two-quantum (2Q) peaks (panel~\textbf{j}). These peaks often indicate bound states of collective excitations, such as biexcitons~\cite{Hu1990,You2015,stone2009two,karaiskaj2010two}, bimagnons~\cite{Wortis1963,Lorenzana1995}, or biphonons~\cite{Cohen1969}. When the resonance linewidth is smaller than the binding energy $\hbar\varepsilon$, the 2Q signal exhibits characteristic splitting along $\omega_t$, with peaks centered at $\omega_t = \omega_0$ and $\omega_t = \omega_0 - \varepsilon$, and a shift along $\omega_\tau$ to $\omega_\tau = 2\omega_0 - \varepsilon$.  For instance, Ref.~\cite{Shahbazyan2000} describes how 2Q memory effects produce a non-exponential temporal profile of four-wave-mixing signals of magnetoexcitons, arising from transitions involving either a bound state or non-perturbative scattering in the dissociated (unbound) exciton-exciton continuum. 
\setcounter{figure}{1}
\pagebreak

\section*{Nonlinear light-matter interactions and collective dynamics in semiconductors}

\begin{figure}[!ht] 
\centering 
\includegraphics[width=\textwidth]{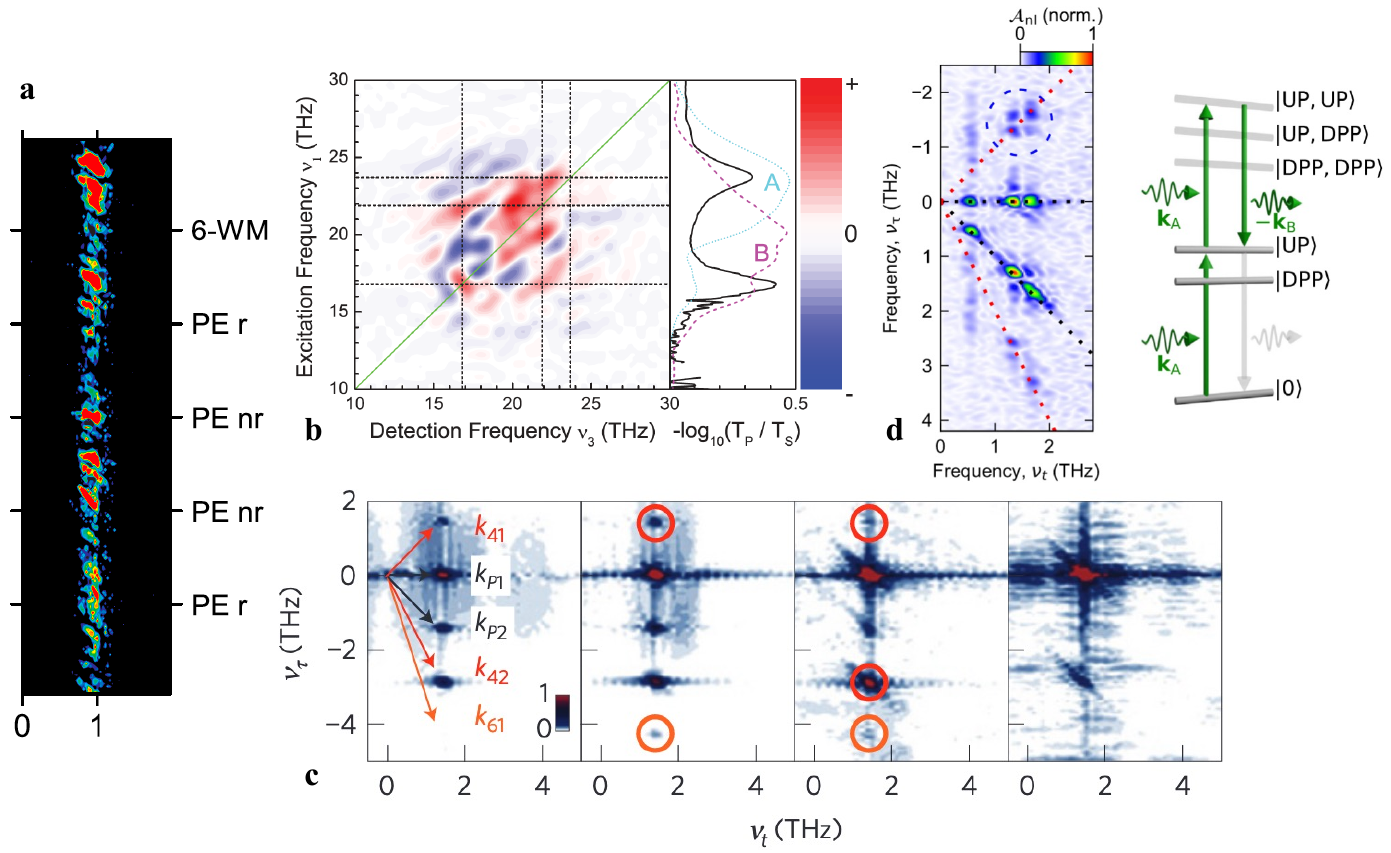} 
\caption{\textbf{Nonlinear light-matter coupling and collective excitations in semiconductors revealed by THz-2DCS.} 
\textbf{a}, 
Intersubband Rabi flopping in GaAs/AlGaAs quantum wells produces multiple THz-2DCS peaks, including  six-wave mixing (6WM), rephasing (r) and  non-rephasing (nr) photon echoes (PE).
\textbf{b}, THz-2DCS spectra of GaAs/AlGaAs double quantum wells reveal strong polaronic effects, with multiple diagonal and off-diagonal peaks arising from interactions between intersubband electronic excitations and longitudinal optical phonons.  
\textbf{c},  THz-2DCS spectra of a two-dimensional electron gas in a static magnetic field for increasing THz field strengths. Signals from pump–probe (PP, black), four-wave mixing (4WM, red), and six-wave mixing (6WM, orange) processes emerge due to strong-field-induced Landau-level dynamics dominated by Coulomb interactions between electrons and the ion background. 
\textbf{d}, Nonlinear response of polariton modes in a custom-designed strongly light–matter-coupled structure. The 2D spectrum shows multi-mode polariton mixing, including four-wave mixing involving virtual energy levels. 
Panel~\textbf{a} adapted from Ref.~\cite{Woerner2013}. Panel~\textbf{b} adapted from Ref.~\cite{Kuehn2011QW2}. Panel~\textbf{c} adapted from Ref.~\cite{Maag2016}. Panel~\textbf{d} adapted from Ref.~\cite{Mornhinweg2021}.
} 
\label{fig_semi}
\end{figure}

By leveraging phase-sensitive detection and multidimensional spectral analysis, THz-2DCS was initially developed as a powerful tool to probe and manipulate ultrafast low-energy electronic and vibrational dynamics, as well as higher-order nonlinear responses in semiconductors~\cite{Kuehn2009,Kuehn2011QW1,Kuehn2011QW2,Junginger2012,Woerner2013,Gill2024,Bowlan2014,Elsaesser2015,Somma2016InSb,Somma2016beyond,Raab2019,Woerner2019,Houver2019,Markmann2020,Raab2020,Tarekegne2020,Mahmood2021,Ghalgaoui2020GaAs,Riepl2021,Woerner2021,Maag2016,Mornhinweg2021}.   

\subsection*{Nonlinear intersubband dynamics and electron-phonon interactions}
 
 Kuehn~\textit{et al.}~\cite{Kuehn2009,Kuehn2011QW1} demonstrated that, in GaAs/AlGaAs multiple quantum wells (QWs), THz-2DCS reveals predominantly homogeneous broadening of intersubband (IS) transitions through free induction decay signatures in four-wave mixing THz-2DCS signals. As the excitation intensity increases, higher-order nonlinearities become significant--up to eight-wave mixing--manifesting as discrete peaks and satellites in the 2D frequency domain.
 At even higher field strengths, the system transitions from a perturbative susceptibility response to a non-perturbative regime, where Rabi oscillations emerge on the IS transition, leading to population inversion between subbands and  photon echo, six-wave mixing, rephasing
and non-rephasing  THz-2DCS peaks (Fig.~\ref{fig_semi}\textbf{a}). This regime, characterized by the breakdown of the rotating wave approximation, was further substantiated by Junginger~\textit{et al.}~\cite{Junginger2012} in bulk InSb using electric field amplitudes exceeding 5~MV/cm. 


In addition to pure electronic transitions, strong coupling between IS transitions and longitudinal optical (LO) phonons has been revealed. Kuehn~\textit{et al.}~\cite{Kuehn2011QW2} demonstrated that IS transitions in GaAs/AlGaAs double QWs exhibit strong coupling to LO phonons, producing pronounced polaronic effects (Fig.~\ref{fig_semi}\textbf{b}). Unlike linear optical spectra,  the 2D spectrum exhibits a multitude of distinct diagonal and off-diagonal multiple peaks that serve as fingerprints of IS  electron-phonon coupling
significantly enhanced as compared to bulk materials. 

\subsection*{Nonlinear Landau-level dynamics}

Among the diverse semiconductor systems explored with THz-2DCS, two-dimensional electron gases (2DEGs) in QWs offer a unique platform to study many-body quantum coherence under strong magnetic fields, where quantized Landau levels govern the electron dynamics. A compelling example of THz-2DCS applied to such systems is provided by Maag~\textit{et al.}~\cite{Maag2016}, who employed collinear THz-2DCS to probe the nonlinear response of a 2DEG in GaAs QWs. Under strong THz excitation, the nonlinear response revealed pronounced anharmonic dynamics, including Landau ladder climbing and coherent multi-wave mixing. The 2D spectra (Fig.~\ref{fig_semi}\textbf{c}) show pump-probe (PP), four-wave mixing (4WM), and six-wave mixing (6WM) peaks, demonstrating the emergence of high-order nonlinearities. These effects are traced primarily to dynamic Coulomb interactions between electron and the positively charged ion background--interactions not accounted for in Kohn's theorem. By accessing internal degrees of freedom in a Landau-quantized system, this study demonstrates how THz-2DCS can unveil many-body nonlinearities even in system traditionally viewed as harmonic and weakly interacting.

In the strong light-matter coupling regime, resonant light fields can drive extreme nonlinearities in hybrid light-matter states such as polaritons. A recent 2DCS experiment performed by Mornhinweg~\textit{et. al.}~\cite{Mornhinweg2021} explored this regime using a tailored structure that couples the eigenmode of a metallic THz resonator to the cyclotron resonance of a 2DEG in six GaAs QWs. This study engineered the cavity Rabi frequency ($\Omega_R^\nu$) and the Landau levels ($h\nu_\text{c}$) to strongly couple with the THz driving field $\omega_\text{c}$, thus giving rise to multiple polariton modes. The nonlinear response of these modes was resolved via THz-2DCS (Fig.~\ref{fig_semi}\textbf{d}, left), revealing complex multi-polariton excitation pathways (right). As shown in Fig.~\ref{fig_semi}\textbf{d}, the  2D spectrum includes dashed lines that trace third-order signals from individual polariton branches. Notably, the experiment also detected coherent inter-polariton mixing, producing additional off-diagonal signals. These interactions signal the breakdown of the normal-mode approximation under strong-field coupling conditions, highlighting the rich nonlinear physics enabled by polaritonic systems.
\section*{Collective modes, echoes, and order parameter dynamics in superconductors}

Ultrafast THz excitation has enabled direct access to the nonlinear dynamics of superconductors, revealing complex interplay among quasiparticles, collective modes, and light-induced symmetry breaking~\cite{Klein1980,Littlewood1981,Podolsky2011,Pekker2015,krull2016,Cea2016,Aoki2017,Wu2019,Schwarz2020,Klein2010,Blumberg,Giorgianni2019,Shimano2019,seibold2021,udina2022thz,Katsumi2024NbN,Katsumi2024MgB2,yuan2024}. These driven phenomena can be broadly classified into three dynamical regimes: (i) an overdamped regime, where superconducting coherence is rapidly quenched by scattering and the system is dominated by quasiparticle pre-thermalization and relaxation~\cite{yang2018}; (ii) a Landau-damped regime, where short-lived Higgs oscillations decay through coupling to the quasiparticle continuum~\cite{Matsunaga:2012,vaswani2019discovery,Luo2023}; and (iii) a regime of persistent coherence, characterized by long-lived collective modes and higher-harmonic generation~\cite{yang2019lightwave}.
A striking feature of the coherent regime is the generation of light-induced DC supercurrents, driven by electromagnetic propagation in thin films. These currents dynamically break inversion symmetry, even in centrosymmetric superconductors, enabling excitation of symmetry-forbidden modes~\cite{yang2019lightwave,vaswani2019discovery}. The resulting nonequilibrium condensate states, characterized by finite-momentum pairing and modified gap dynamics, offer new avenues for manipulating the superconducting order and exploring emergent collective phenomena on ultrafast timescales.

THz-2DCS has become a powerful technique for probing these light-driven states beyond the capabilities of conventional linear and pump-probe spectroscopies~\cite{Matsunaga:2012,Matsunaga:2013,matsunaga2014,Matsunaga2017,Giorgianni2019,Yang2019b,Chu2020,hybrid-higgs,cheng2023lowenergy}. By resolving phase-sensitive higher-order nonlinear responses across excitation and detection frequencies, THz-2DCS disentangles overlapping quasiparticle and collective-mode signals, separate homogeneous and inhomogeneous broadening, and reveals quantum pathways. This capability has been leveraged to study Higgs and Leggett modes~\cite{Katsumi2024NbN,Katsumi2024MgB2}, light-induced Floquet-like states~\cite{Luo2023}, nonlinear order parameter dynamics~\cite{Kim2024}, and quantum echoes~\cite{Huang2023,Liu2024}. These experiments are supported by complementary theoretical frameworks, including disorder-based models~\cite{Salvador2024,salvador2025}, diagrammatic approaches~\cite{Katsumi2024NbN}, pseudospin dynamics~\cite{Manske2023,Puviani2024}, and gauge-invariant Maxwell–Bloch simulations~\cite{Mootz2020,Mootz2022,mootz2023multidimensional}, which illuminate the underlying dynamics. 
These advancements place THz-2DCS at the forefront of nonlinear optics in superconductors, establishing it as a unique platform for exploring coherent dynamics in quantum materials and uncovering novel mechanisms of quantum coherence, symmetry breaking, and collective excitations that shape the nonequilibrium high-$T_\text{c}$ superconductivity and light-induced superconducting states~\cite{Zhou2021,Fausti2011,Mitrano2016,Budden2021,Rowe2023,Chattopadhyay2025}.

\begin{figure}[!ht] 
\centering 
\includegraphics[width=\textwidth]{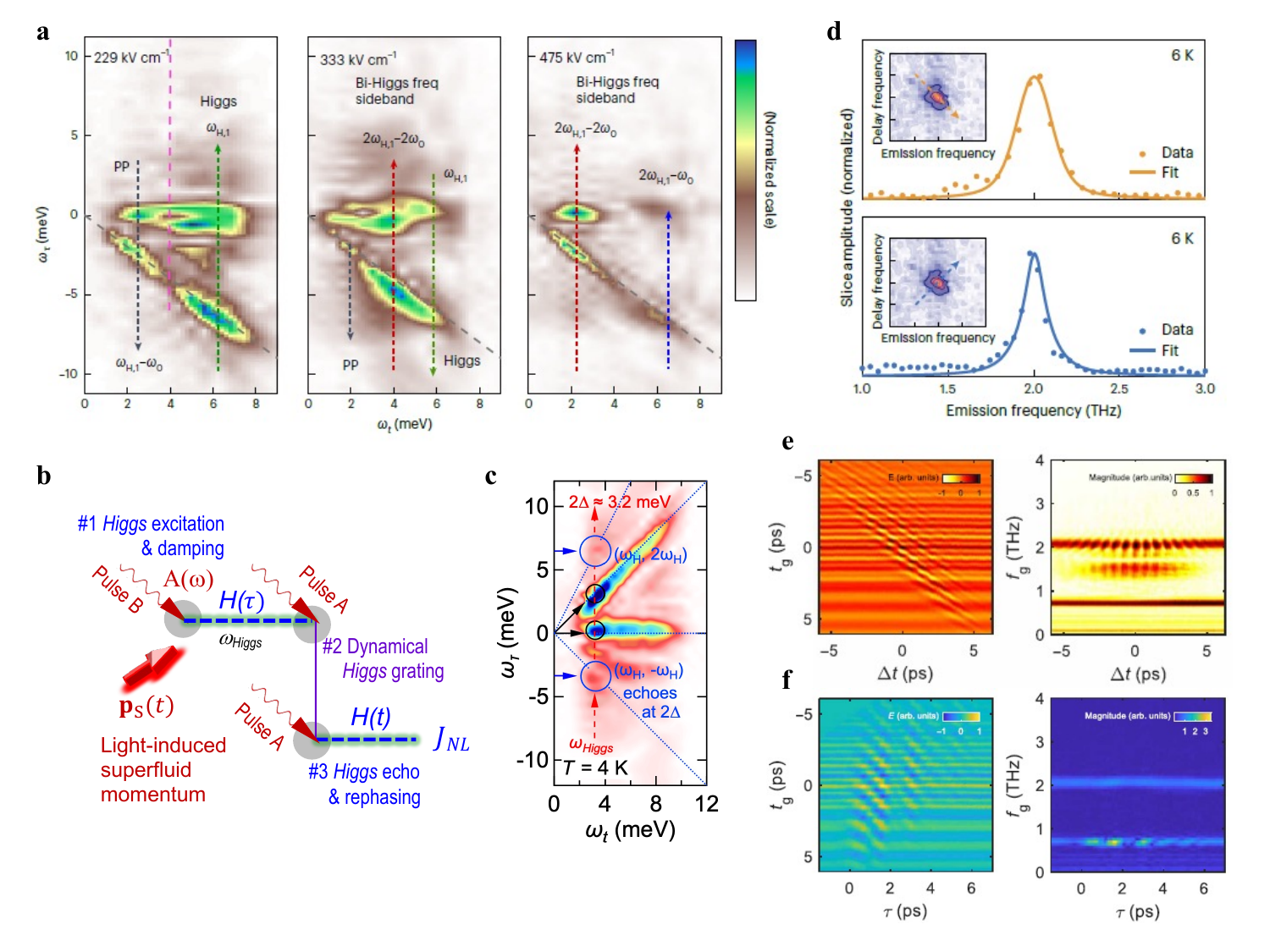} 
\caption{\textbf{Collective modes, echoes, and order parameter dynamics in superconductors revealed by THz-2DCS.} 
\textbf{a}, THz-2DCS of an iron-based superconductor. Experimental 2D spectra at pump fields of 229~kV/cm, 333~kV/cm, and 475~kV/cm show a transition from perturbative Higgs peaks (green dashed line) to dominant bi-Higgs sidebands (red and blue dashed lines) above a critical field threshold.  
\textbf{b}, Higgs-echo spectroscopy of a Nb superconductor. Schematic of the three-step Higgs echo process: (1) THz-driven Higgs excitation $H(t)$ via inversion symmetry breaking induced by a persisting superfluid momentum $\mathbf{p}_\mathrm{S}(t)$, followed by Higgs dephasing over delay $\tau$; (2) formation of a temporal Higgs grating by interaction with second pulse; (3) echo generation via scattering of quasiparticle coherence off this grating. $J_\mathrm{NL}$: nonlinear current responsible for THz emission; $A$: vector potential of THz light fields. 
\textbf{c}, 2D Fourier spectrum of the Nb nonlinear signal at 4~K, showing Higgs pump--probe peaks at $(\omega_\mathrm{H}, 0)$ and $(\omega_\mathrm{H}, \omega_\mathrm{H})$ (black circles), and echo peaks at $(\omega_\mathrm{H}, -\omega_\mathrm{H})$ and $(\omega_\mathrm{H}, 2\omega_\mathrm{H})$ (blue circles). 
\textbf{d}, Josephson echo spectroscopy of LSCO. Temperature-dependent linewidth analysis of Josephson plasmon peak in LSCO reveals disorder-induced broadening. Insets: Anisotropic Josephson plasmon peak near $(2,-2)$~THz. Line cuts along anti-diagonal (orange) and diagonal (blue) directions reflect contributions from intrinsic and disorder effects.
\textbf{e}, \textbf{f} Quench-drive spectroscopy combining an optical quench pulse with a delayed multicycle THz field $E_\text{THz}(\Omega = 0.7$~meV) in LSCO ($T_\text{c} = 44$~K, $\Delta_\text{sc} \sim 20$~meV). \textbf{e}, Left: Differential THz signal vs. gate time $t_\text{g}$ and quench-drive delay $\Delta t$. Right: Transient spectrum after Fourier transform along $t_\text{g}$ reveals FH and THG peaks at 0.7 and 2.1~THz, and a 1.4~THz sideband from four- and six-wave mixing between THz and optical fields. 
\textbf{f}, Left: Sheared analysis at constant $\tau = t_\text{g} + \Delta t$ suppresses quasiparticle screening at THG, isolating the intrinsic superconducting response. Right: Corresponding transient spectrum. 
Panel~\textbf{a} adapted from Ref.~\cite{Luo2023}.
Panels~\textbf{b}--\textbf{c} adapted from Ref.~\cite{Huang2023}.
Panel~\textbf{d} adapted from Ref.~\cite{Liu2024}. 
Panels \textbf{e}--\textbf{f} adapted from Ref.~\cite{Kim2024}.
} 
\label{fig_sc}
\end{figure}

\subsection*{Characterizing driven superconducting states through their collective modes}

THz-2DCS enables spectrally-resolved tracking of nonequilibrium superconducting dynamics, identifying amplitude, phase, and hybrid collective modes arising from light-induced symmetry breaking and strong correlations. 
In iron-based superconductors (FeSCs) hosting coupled Higgs collective modes~\cite{hybrid-higgs}, Luo~\textit{et al.}~\cite{Luo2023} demonstrated that intense phase-locked THz fields drive nonlinear coupling between amplitude and phase fluctuations of the SC order parameters in different bands. Above a critical field, this leads to a dynamical phase-amplitude mode with excitation energy set by the Higgs frequency $\omega_\mathrm{H}$.
THz-2DCS spectra (Fig.~\ref{fig_sc}\textbf{a}) show a transition from conventional hole-pocket Higgs mode peaks at $\omega_t = \omega_{\text{H},1} \approx 6.8$~meV 
to bi-Higgs sidebands at $2\omega_{\text{H},1} - 2\omega_0$ and $2\omega_{\text{H},1} - \omega_0$, where  $\omega_0 \approx 4$~meV is the THz pulse center frequency. Quantum kinetic simulations~\cite{Mootz2020,Mootz2022,mootz2023multidimensional} attribute this to a Floquet-like SC state with time-periodic amplitude-phase collective mode dynamics. At lower fields, the spectra reveal distinct peaks directly linked to the SC energy gap, appearing at characteristic 2D frequency coordinates such as $(\omega_{\text{H},1}, 0)$ and $(\omega_{\text{H},1}, \omega_{\text{H},1})$ (Fig.~\ref{fig_sc}\textbf{b}, left panel), where no signal is expected without light-induced symmetry breaking or disorder. These are attributed to resonantly excited Higgs modes sustained by THz-induced DC supercurrents.
The technique’s sensitivity to detecting nonequilibrium states and their collective modes has been further demonstrated across a range of SC systems--resolving the Higgs mode in NbN~\cite{Katsumi2024NbN} and Nb~\cite{Huang2023} superconductors, detecting both Higgs and Leggett modes in multiband MgB$_2$ superconductors~\cite{Katsumi2024MgB2,yuan2024}, and probing  cuprate-like nonlinearities in SC infinite-layer nickelates~\cite{Cheng2025}. These studies highlight the technique’s ability to uncover complex collective behavior in diverse superconductors.

 
\subsection*{Echo spectroscopy}

The recent development of Higgs echo spectroscopy has introduced a powerful strategy for probing coherent dynamics of collective modes in superconductors. Using THz-2DCS, Huang~\textit{et al.}~\cite{Huang2023} reported the first observation of photon echo signals at the Higgs mode frequency in a niobium thin film. These echoes reveal rephasing of long-lived Higgs oscillations and provide direct evidence of coherent memory effects in nonequilibrium SC states--phenomena inaccessible to conventional time-domain or third-order techniques. The mechanism, illustrated in Fig.~\ref{fig_sc}\textbf{b}, involves a three-step quantum pathway. A first THz pulse excites the Higgs mode via light-induced supercurrents that break inversion symmetry, enabling linear coupling to the light field. A second, phase-coherent pulse modulates the order parameter, creating a transient “Higgs grating”. A subsequent quasiparticle coherence then scatters off this grating, producing a delayed echo. The measured THz-2DCS spectrum (Fig.~\ref{fig_sc}\textbf{c}) shows echo peaks at $(\omega_t, \omega_\tau) = (\omega_\mathrm{H}, - \omega_\mathrm{H})$ and $(\omega_\mathrm{H}, 2\omega_\mathrm{H})$ (blue circles), alongside Higgs pump-probe peaks at $(\omega_t, \omega_\tau) = (\omega_\mathrm{H}, 0)$ and $(\omega_\mathrm{H}, \omega_\mathrm{H})$ (black circles). The echo is characterized by an asymmetric temporal profile and negative-time signals, indicating anharmonic coupling between Higgs and quasiparticle excitations. Quantum kinetic simulations based on gauge-invariant superconductor Bloch equations attribute the echo to interference between two phase-locked Higgs excitations, which generates a dynamical Higgs grating that scatters quasiparticle coherence. These results demonstrate that THz-2DCS effectively uncovers complex dynamical processes, including Higgs rephasing, nonlinear coupling, and symmetry breaking from highly quantum-entangled states.
Combined with advanced time-frequency analysis~\cite{Huang2023}, these findings differentiate Higgs from quasiparticle responses and identify asymmetric echo delays arising from inhomogeneous broadening and dynamically evolving ``soft" quasiparticle bands under THz excitation.

Echo-based 2D spectroscopy has also proven effective for probing interlayer coherence in cuprate superconductors. Liu~\textit{et al.}~\cite{Liu2024,Salvador2024} explored interlayer Josephson echoes originating from coherent tunneling between superconducting layers in near-optimally doped La$_{1.83}$Sr$_{0.17}$CuO$_4$ (LSCO) superconductors, using a non-collinear phase-matching geometry to isolate momentum-dependent components of the third-order susceptibility $\chi^{(3)}$. The resulting 2D spectrum (insets Fig.~\ref{fig_sc}\textbf{d}) exhibits a sharp echo peak, providing a direct signature of coherent Josephson tunneling. Figure~\ref{fig_sc}\textbf{d} illustrates how spatial disorder modifies this echo response. Anisotropic broadening of the peak along the diagonal and anti-diagonal directions (top and middle panels) reflects the interplay between intrinsic damping and inhomogeneous disorder. These results underscore the unique power of echo-based THz-2DCS to disentangle microscopic decoherence mechanisms in correlated materials. 




\subsection*{Quench-drive spectroscopy}

Quench-drive spectroscopy~\cite{Manske2023,Puviani2024} offers an alternative to conventional THz-2DCS for probing superconducting order parameter dynamics, collective modes, and quasiparticle excitations. Here, a strong optical pulse first quenches the superconducting condensate, followed by a delayed multicycle THz field that drives the system. This configuration enables real-time tracking of the order parameter and its nonlinear optical response. 
Applied to LSCO superconductors~\cite{Kim2024}, this method revealed modulations in both the fundamental harmonic (FH) and third-harmonic generation (THG) spectral peaks, along with emerging sidebands. Two complementary analyses were employed. 
First, by fixing the quench-drive delay $\Delta t$ and scanning the gate time $t_\text{g}$ (Fig.~\ref{fig_sc}\textbf{e}, left), interference patterns between FH and THG components were observed, yielding spectral peaks at $f_\text{FH} = 0.7$~THz and $f_\text{TH} = 2.1$~THz in the frequency domain (right). Additional sidebands at 1.4~THz arise from four- and six-wave mixing processes between the optical and THz fields.
Second, a shearing transformation was applied by analyzing the response at fixed acquisition time $\tau = t_\text{g} + \Delta t$ (Fig.~\ref{fig_sc}\textbf{f}, left) to disentangle the quasiparticle response from the intrinsic dynamics of the SC order parameter. This method effectively suppresses screening from photoexcited quasiparticles at the THG frequency. The corresponding Fourier-transformed spectrum (right) shows that FH amplitude modulations span frequencies from 0.7 to 1.4~THz, while the THG component remains nearly static. From this, the time-dependent superconducting gap $\Delta(\tau)$ is extracted by evaluating the transient nonlinear susceptibility $\chi(\tau)$, which is proportional to the THG amplitude normalized by FH. The resulting dynamics reveals a rapid gap suppression within $\sim 1$~ps, followed by recovery over a $\sim$ 9~ps timescale, thus providing a real-time view of order parameter evolution under strong optical perturbation.
\section*{Nonlinear magnonics and spin correlations}

Probing light-induced nonequilibrium spin dynamics in magnetic systems remains challenging due to the inherently weak magnon-photon coupling~\cite{Kirilyuk2010,Kampfrath2011,Baierl2016,Afanasiev2021,Behovits2023}. Recent advances overcome this limitation by utilizing intense THz magnetic field pulse pairs to resonantly drive coherent spin dynamics~\cite{Pashkin2013,Lu2017,Mashkovich2021,Blank2023Spin, Grishunin2023,Zhang2024Coup,Zhang2024Down,Zhang2024Up,Huang2024,Leenders2024,Zhang2024Spin,Dutta2025}. This two-pulse coherent excitation scheme induces collective spin precession, which appears as macroscopic oscillations of antiferromagnetic (AFM) and ferromagnetic (FM) modes. The resulting long-lived spin precession creates a broad temporal window for coherent optical control via a delayed second light pulse, enabling the disentanglement of coherent interactions and nonlinear coupling between magnons and other quasiparticles~\cite{Chovan2006, Chovan2008, Wang2009, Kapetanakis2009, Lingos2015,Patz2015}. Recent THz-2DCS experiments have successfully implemented this approach and revealed a range of nonlinear magnonic phenomena, including nonlinear spin-wave mixing~\cite{Lu2017,Huang2024, Dutta2025}, magnon-magnon~\cite{Zhang2024Coup,Blank2023Spin} and magnon-phonon~\cite{Mashkovich2021,Grishunin2023} couplings, as well as ultrafast magnon conversion~\cite{Zhang2024Down,Zhang2024Up,Leenders2024}.

\begin{figure}[ht!]
\centering 
\includegraphics[width=\textwidth]{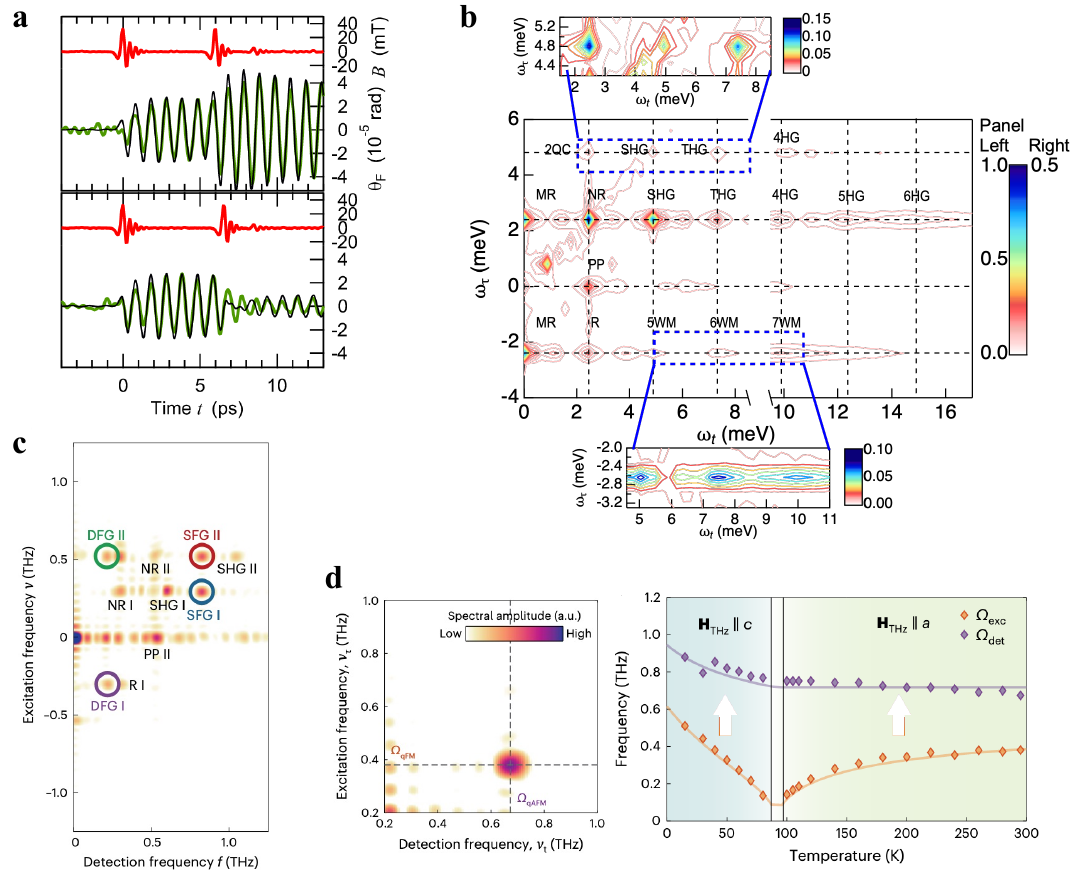} 
\caption{\textbf{THz-2DCS of magnetic nonlinear dynamics.} 
\textbf{a}, Double-pulse excitation (red curves) of NiO. The green curves represent the pump-induced Faraday rotation $\theta_\text{F}$ as a function of time $t$. The second THz pulse is applied either in-phase (top) or out-of-phase (bottom) with respect to the spin precession initiated by the first pulse.
\textbf{b}, THz-induced magnonic multiplication in Sm$_{0.4}$Er$_{0.6}$FeO$_3$. High-order nonlinear signals include four-, five-, six-harmonic generations (4HG, 5HG, 6HG), as well as five-, six- and seven-wave mixing (5WM, 6WM and 7WM) processes. 
\textbf{c}, Light-induced nonlinear magnon-magnon coupling in YFeO$_3$. Coherent mixing between quasi-AFM and quasi-FM modes leads to four additional mode-coupling peaks (sum-frequency generation (SFG) I \& II, difference-frequency generation (DFG) I \& II). 
\textbf{d}, Ultrafast nonlinear magnon upconversion in ErFeO$_3$. A low-frequency quasi-FM mode is upconverted to a high-frequency quasi-AFM mode. Left: 2D spectrum obtained via a cross-polarized detection scheme. Right: Temperature-dependent magnon upconversion across the spin-reorientation transition.
Panel \textbf{a} adapted from Ref.~\cite{Pashkin2013}. 
Panel \textbf{b} adapted from Ref.~\cite{Huang2024}. 
Panel \textbf{c} adapted  from Ref.~\cite{Zhang2024Coup}. 
Panel \textbf{d} adapted from Ref.~\cite{Zhang2024Up}.} 
\label{fig_mag}
\end{figure}


The protocol of THz 2D magnetic  spectroscopy follows conventional THz-2DCS. 
The magnetic fields of the THz pulse pair can be aligned either parallel or perpendicular to the net magnetization $\mathbf{M}$ to coherently excite chosen magnon modes. The emitted nonlinear signals relating to coherent spin dynamics can be detected via electro-optic sampling, enabling direct access to the higher-order nonlinear magnetic responses. Alternatively, a weak near-infrared (NIR) probe pulse can be also applied after the two pump pulses to perform a magneto-optical Kerr effect (MOKE)-type measurement, also referred to as double-pump MOKE spectroscopy. 
This scheme provides insight into the nonequilibrium interactions and energy transfers among photons, magnons, and phonons.

\subsection*{Higher-order magnon  nonlinearities}
The prototype of coherent THz magnon control was first demonstrated by Pashkin~\textit{et al.}~\cite{Pashkin2013} in the antiferromagnet NiO. As shown in Fig.~\ref{fig_mag}\textbf{a}, two phase-locked THz pulses--either in-phase or out-of-phase--coherently enhance or suppress the magnetization oscillations, respectively. Building on this concept, Lu~\textit{et al.}~\cite{Lu2017} reported the comprehensive THz-2DCS study of rare-earth orthoferrite YFeO$_{3}$ (YFO). The measured 2D spectrum reveals a range of $\chi^{(3)}$ nonlinear magnon signals, such as phase-reversed (R, spin echo),  and two-quantum (2Q) coherence peaks that involve in multiple field-spin interactions, as well as $\chi^{(2)}$ nonlinear signals--THz rectification (TR) and second-harmonic generation (SHG)--due to Dzyaloshinskii-Moriya symmetry breaking. Going beyond third-order nonlinearity, Huang~\textit{et al.}~\cite{Huang2024} resolved extreme terahertz magnon multiplication recently, as shown in Fig.~\ref{fig_mag}\textbf{b}. These high order nonlinear magnon interactions, up to six-magnon quanta harmonic generation and seven-wave mixing, reveal the presence of robust spin coherence and underline the role of intrinsic quantum spin effects~\cite{mootz2023twodimensional}.

\subsection*{Nonlinear magnon couplings and upconversion}

In studies of magnon mode couplings, linear response typically dominates the measured  signals, obscuring the underlying nonlinear mechanism. A 
recent breakthrough by Zhang \textit{et al.}~\cite{Zhang2024Coup} has revealed anharmonic magnon-magnon coupling using THz-2DCS for the first time. In their experiment, two THz pulses were aligned parallel to the bisector of  $ac$-axis in YFO, simultaneously driving both quasi-FM and quasi-AFM modes. The nonlinear interaction between these  modes generated the emergence of sum-frequency generation (SFG) and difference-frequency generation (DFG) signals, which were clearly resolved in the 2D spectrum (Fig.~\ref{fig_mag}\textbf{c}). These four additional peaks were attributed to the second order magnetic process, where each magnon mode interacts once with the THz fields. For instance, the SFG peaks at ($\Omega_{\mathrm{det}}=\Omega_\mathrm{qAFM}+\Omega_\mathrm{qFM}, \ \Omega_\mathrm{exc}=\Omega_\mathrm{qAFM}$ and $\Omega_\mathrm{qAFM}$) demonstrate two distinct excitation pathways, determined by the sequence in which the modes are driven first. Such second-order nonlinear responses 
confirm the existence of coherent magnon–magnon mixing, exhibiting a unique anisotropic coupling between qFM and qAFM modes.

One promising application of nonlinear magnon control is the ultrafast magnon conversion. Nonequilibrium magnon states induced by the first pump pulse can significantly enhance their coupling to external fields (second pulse), enabling the inter-magnon conversion process. Orthoferrites with canted spin orders provide an ideal platform for realizing this effect~\cite{Leenders2024}. As shown in Fig.~\ref{fig_mag}\textbf{d}, ultrafast magnon upconversion was demonstrated in ErFeO$_3$ by using THz-2DCS~\cite{Zhang2024Up}. Unlike previous mentioned SFG/DFG experiment (Fig.~\ref{fig_mag}\textbf{c}), a cross-polarized detection scheme was applied to  reveal an intrinsic upconversion from the lower energy qFM mode to the higher energy qAFM mode, resulting a distinct spectral peak at $(\Omega_\mathrm{qAFM},\Omega_\mathrm{qFM})$ (left panel, Fig.~\ref{fig_mag}\textbf{d}).  The absence of a downconversion 2D spectral peak at $(\Omega_\mathrm{qFM},\Omega_\mathrm{qAFM})$ indicates an unidirectional energy transfer channel in the system.   
The temperature-dependent measurements (right panel, Fig.~\ref{fig_mag}\textbf{d}) further confirm the robustness of this magnon upconversion process. 
Notably, despite the net magnetization $\mathbf{M}$ rotating from the $c$-axis to the $a$-axis due to a spin reorientation transition, the upconversion  persists over a broad temperature range, provided the THz excitation field $\mathbf{H}_\mathrm{THz}$ remains perpendicular to $\mathbf{M}$.

\section*{Other emerging excitations uncovered by THz-2DCS}

Beyond semiconductors, superconductors, and magnets reviewed so far, THz-2DCS is rapidly expanding into a broader range of research areas \cite{Blank2023Phonon,Bhandia2024,Barbalas2025,Somma2014,Lin2022,Folpini2017,Pal2021,Lu2016,Zhang2021,Allodi2015,Finneran2016,Finneran2017,Magdau2019,Mead2020,Runge2023Sol, Ghalgaoui2020Water,Runge2023Bis, Biggs2023,Johnson2019,Reimann2021}, advancing our understanding of complex light-matter interactions through nonlinear responses driven by lattice motions and its softness~\cite{Folpini2017,Pal2021,Johnson2019,Biggs2023,Runge2023Bis}, molecular rotations~\cite{Lu2016,Zhang2021}, and molecular vibrational transitions~\cite{Allodi2015,Finneran2016,Finneran2017,Magdau2019,Mead2020}. The exemplified topics presented in this section not only illuminate fundamental quantum phenomena, but also highlight the growing role of THz-2DCS in characterizing quantum materials with unconventional properties. These advances establish THz-2DCS as a powerful platform for probing nonlinear quantum dynamics across diverse material systems. Its ability to capture the interplays among electronic, lattice, spin, and photonic degrees of freedom open new opportunities for controlling quantum coherence and energy transport. Future developments promise to bridge fundamental discoveries with technological innovation, driving breakthroughs in quantum engineering, photonics, and the design of functional quantum materials.

\begin{figure}[!ht] 
\centering 
\includegraphics[width=\textwidth]{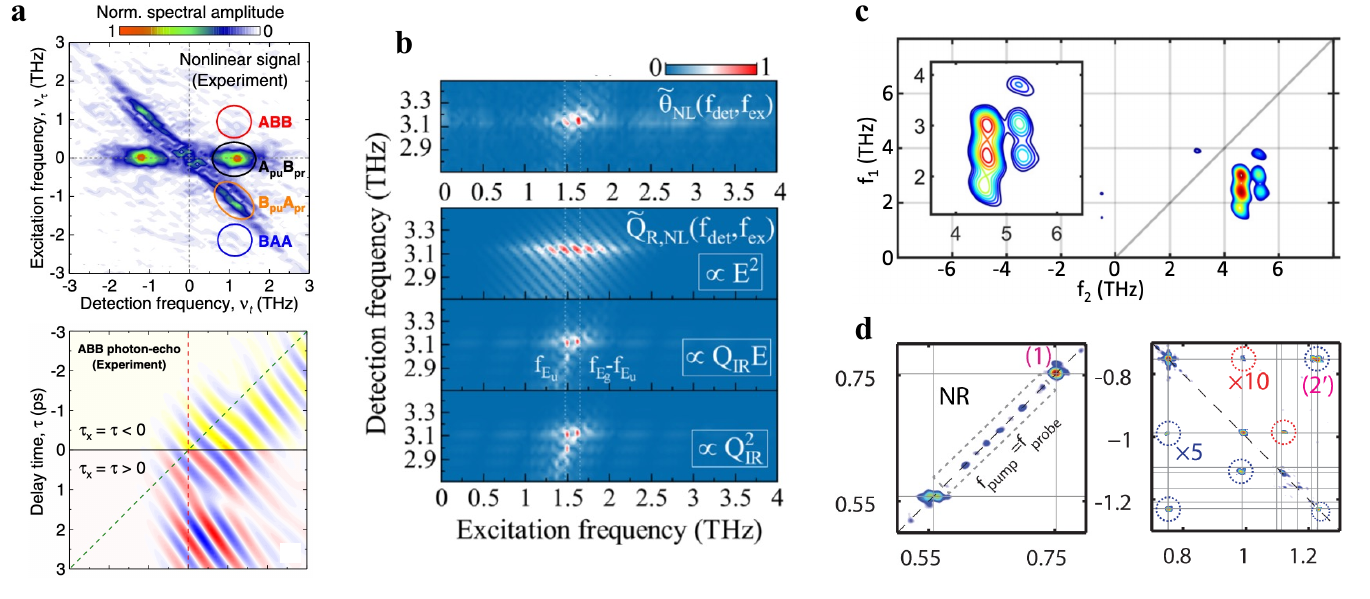} 
\caption{\textbf{Other emerging topics in THz-2DCS}. 
\textbf{a}, Soft-mode nonlinearities in the ferroelectric perovskite SrTiO$_3$. Top: 2D spectrum of the soft-mode nonlinear response, where ABB, A$_\mathrm{pu}$B$_\mathrm{pr}$, B$_\mathrm{pu}$A$_\mathrm{pr}$, and BAA peaks represent $\chi^{(3)}$ signals from the soft mode at $\nu_0 = 1.17$~THz. Bottom: Inverse Fourier transform of the ABB echo peak. Positive $\tau$ signals arise from conventional photon-echo coherence, while negative $\tau$ signals reflect non-instantaneous local-field effects. 
\textbf{b}, Raman-active $E_\text{g}$ phonon excitation via a nonlinear photophononic process in MnBi$_2$Te$_4$. Nonlinear polarization signals compared with simulations of three possible driving mechanisms: purely photonic ($E^2$), hybrid photophononic ($Q_\mathrm{IR}E$), and purely phononic ($Q_\mathrm{IR}^2$). 
\textbf{c}, Representative 2D terahertz-terahertz-Raman spectrum of liquid molecule CBr$_2$Cl$_2$, with $f_1$ and $f_2$ axes corresponding to the THz pump and optical probe delays, respectively.
\textbf{d}, Nonlinear rotational spectroscopy in water vapor at 60$^\circ$C. Horizontal and vertical axes represent detection ($f_\mathrm{probe}$) and excitation ($f_\mathrm{pump}$) frequencies. Dashed circles highlight off-diagonal peaks involving two-quantum (2Q) coherences.
\textbf{a} adapted from Ref.~\cite{Pal2021}.
\textbf{b} adapted from Ref.~\cite{Blank2023Phonon}. 
\textbf{c} adapted from Ref.~\cite{Finneran2016}.
\textbf{d} adapted from  Ref.~\cite{Zhang2021}.
} 
\label{fig_others}
\end{figure}

\subsection*{Soft-mode nonlinearities}

Using 2D THz spectroscopy, Pal \textit{et al.}~\cite{Pal2021} investigated the origin of soft-mode nonlinearities in the ferroelectric perovskite SrTiO$_3$, uncovering strong coupling between electronic interband transitions and lattice dynamics. The experimental approach followed standard THz-2DCS protocols, utilizing GaP and ZnTe crystals as THz emitters for the first and second excitation pulses, respectively. Figure~\ref{fig_others}\textbf{a} presents contour plots of the nonlinear spectrum (top panel) and ABB photon-echo dynamics (bottom panel). The 2D spectrum displays two dominant pump-probe peaks (A$_\mathrm{pu}$B$_\mathrm{pr}$ and B$_\mathrm{pu}$A$_\mathrm{pr}$) and two weaker photon-echo peaks (ABB and BAA), all centered at the soft-mode frequency $\nu_0 = 1.17$~THz. The broadened lineshapes of the pump-probe peaks indicate a field-driven blueshift of the soft-mode resonance. Further analysis of the AAB peak using inverse Fourier transform reveals lasting oscillations at both negative and positive coherent times ($\tau<0$ and $\tau>0$), pointing to persistent local-field effects mediated by electronic and ionic dipole interactions. These findings support a hybrid lattice-electronic picture of soft-mode dynamics in SrTiO$_3$, and provide a conceptual framework for exploring softness in other complex materials. 

\subsection*{Nonlinear photophononic processes}
Manipulating lattice dynamics presents a promising strategy for controlling topological quantum states. In a study of nonlinear phononics in the topological antiferromagnet MnBi$_2$Te$_4$, Blank \textit{et al.}~\cite{Blank2023Phonon} employed THz-2DCS to map energy transfer pathways involving nonlinear excitation of the Raman-active $E_\text{g}$ phonon mode. By monitoring polarization changes of a weak near-infrared (NIR) probe pulse as functions of THz pulse-pair delay $\tau$ and probe delay $t$, they resolved coherent phonon dynamics with femtosecond time resolution. Figure~\ref{fig_others}\textbf{b} shows the 2D Fourier transform of the measured nonlinear signal, revealing a dominant peak at $(f_\mathrm{exc} = 1.5~\mathrm{THz}, f_\mathrm{det} = 3.14~\mathrm{THz})$, indicating that the $E_\text{g}$ phonon is excited via a 1.5~THz intermediate state. Theoretical modeling ruled out purely photonic or phononic pathways and identified a nonlinear photophononic mechanism: the first THz pulse coherently drives an IR-active $E_\text{u}$ phonon, which then interacts with the second pulse to populate the final $E_\text{g}$ Raman-active state. This work resolves longstanding debates surrounding the excitation of Raman-active phonons in Bi$_2$(Se,Te)$_3$-based materials and provides a conceptual framework for understanding nonlinear phononics in topological quantum materials.

\subsection*{Molecular nonlinear response}
Nonlinear THz spectroscopy offers another powerful tool to probe molecular coherences and dynamics on the meV scale, where many vibrational and rotational transitions occur. Two experimental approaches have been developed to investigate such processes. The first, known as two-dimensional terahertz-terahertz-Raman (2D-TTR) spectroscopy, was introduced by Finneran \textit{et al.}~\cite{Finneran2016}. This technique employs cross-polarized THz pump pulses followed by a NIR probe to detect birefringence changes induced by coherent molecular vibrations. For instance, in liquid CBr$_2$Cl$_2$, 2D-TTR spectroscopy revealed six spectral peaks resulting from anharmonic coupling between multiple vibrational modes (Fig.~\ref{fig_others}\textbf{c}), offering quantitative insights into intermode coupling strengths and vibrational energy transfer pathways.

In a complementary approach, Zhang \textit{et al.}~\cite{Zhang2021} employed conventional THz-2DCS to study nonlinear rotational transitions in water vapor (Fig.~\ref{fig_others}\textbf{d}). The resulting 2D spectra exhibit dominant non-rephasing (NR) and rephasing (R, echo) signals originating from coherent rotational dynamics of ortho-H$_2$O and para-H$_2$O species. For example, diagonal NR peaks at 0.558~THz and 0.753~THz correspond to the $1_{01} \rightarrow 1_{10}$ transition of ortho-H$_2$O and the $2_{02} \rightarrow 2_{11}$ transition of para-H$_2$O, respectively. Notably, the absence of anti-diagonal peaks at (0.558 THz, 0.753 THz) or (0.753 THz, 0.558 THz) signifies forbidden energy transfer between ortho- and para-H$_2$O species. In contrast, the presence of cross-peaks (dashed circles, Fig.~\ref{fig_others}\textbf{d}) arises from coherences between energy levels within the same molecular species. These results demonstrate how THz-2DCS can be used to resolve and distinguish molecular species based on their nonlinear rotational correlations, providing a powerful tool for disentangling complex molecular dynamics.

\section*{Outlook}

The advancement of THz-2DCS will critically depend on the ability to reconstruct quantum many-body states and disentangle excitation pathways under extreme tuning conditions and scales. Such measurements are highly desirable for probing multi-order correlation functions and collective modes under high pressures~\cite{Mao2018} and strong magnetic fields~\cite{kono}. Furthermore, achieving and understanding quantum functionalities demands coherent control of light-matter interactions and dynamics at the quantum limit--pushing the boundaries of space-time-frequency resolution, particularly at nanometer–femtosecond–terahertz (nm–fs–THz) scales~\cite{kim2023visualizing,Tokura2017,Huebener2021}. We identify two critical near-term priorities to propel THz-2DCS forward, enabling new frontiers in quantum materials and technology research.


\subsection*{Advancing THz nonlinear spectroscopy under extreme conditions and scales}
These additional dimensions act as powerful tuning parameters, granting access to diverse phase spaces, modifying the initial conditions of correlated electronic states, and revealing hidden quantum phenomena~\cite{Yamamoto2015, Sebastian2012}. 
Realizing THz-2DCS experiments under such extreme conditions and scales requires advanced instrumentation, including persistent or pulsed cryogenic magnets for strong magnetic field generation, single-shot THz electro-optic sampling techniques, near-field microscopy, and diamond anvil cells (DACs) with carefully tailored culet sizes to achieve ultra-high pressures, as detailed below:

\begin{itemize}

\item \textit{\textbf{Magneto-THz 2DCS:}}
This framework utilizes not only the oscillatory electric fields of THz laser pulse pairs, but also their precise phase control under strong magnetic field conditions--up to 10~T in static fields \cite{Maag2016} and as high as tens of Tesla in pulsed geometries \cite{kono}. 
Intense magnetic fields can 
substantially modify the quantum geometry, topology, and electronic correlations in a material, 
thereby offering deeper insights into nontrivial collective responses and emergent quantum states that are inaccessible at zero field.

\item \textit{\textbf{THz-2DCS under high pressure:}} High-pressure conditions provide a powerful additional dimension for exploring quantum materials, as modifying interatomic distances can induce dramatic changes in electronic structure, spin configurations, and lattice dynamics, as demonstrated by recent advances in ultrafast spectroscopy under high pressure~\cite{McMillan2002,Zhou2016,Dias2017,Mao2018}. Integrating diamond anvil cells (DACs) into THz-2DCS setups will enable to 
track correlation functions and collective modes at high pressures up to 100~GPa depending on the sample size. 
This new capability enables the reconstruction of phase diagrams and energy landscapes that are inaccessible at ambient conditions. 
When combined with magnetic fields and low temperatures \cite{JW}, pressure-tuned THz-2DCS spectra may offer a unique vantage point for unraveling competing quantum orders and novel phases. 

\item \textit{\textbf{THz 2D coherent microscopy (THz-2DCM):}} With the advent of scattering-type scanning near-field optical microscopy (sSNOM), researchers can now confine intense THz pulses onto a sharp metallic tip in an atomic force microscope (AFM), achieving sub-20~nm spatial resolution by inducing localized oscillations of polarization charges or free carriers on the sample surface~\cite{kim2021terahertz,kim2022terahertz,kim2023visualizing,Kim2024, Zhang2018,Guo2024}. 
Near-field signals are detected through phase-resolved EOS of the scattered THz radiation--a method fully compatible with THz-2DCS detection. 
Building on these capabilities, a key next step is to integrate THz-2DCS into this near-field platform to create THz-2DCM, 
enabling sub-wavelength spatial resolution of ultrafast THz signals.

\item \textit{\textbf{Cryogenic-magneto-THz 2DCM:}} Exploring many fascinating quantum behaviors and correlation phenomena in Table~\ref{table} normally requires magnetic and liquid helium cryogenic conditions.  
Recently, an emerging experimental progress brings THz-sSNOM into extreme environments, enabling measurements at temperatures below 2~K and magnetic fields exceeding 5~T~\cite{kim2023sub}. This technical progress paves a way for integrating such cutting-edge instrument with THz-2DCS to develop the cryogenic-magneto THz-2DCM, aiming to extend THz-2DCS capabilities into high-field, low-temperature, and nanoscale spatial resolution regimes. It establishes one of the most compelling frontiers in ultrafast spectroscopy, nano-optics, and quantum materials research.

\end{itemize}

\subsection*{Emerging scientific directions in THz-2DCS: Exotic coherences and quantum technologies}

Throughout this Technical Review, THz-2DCS has demonstrated exceptional sensitivity to collective modes such as Higgs modes, magnons, phonons and Landau fermions.
Further integration of experimental and theoretical frameworks in THz-2DCS is crucial for uncovering exotic collective modes and elucidating their underlying mechanisms--particularly in coherent, transient, and metastable states of superconducting, topological, and magnetic quantum materials, as detailed below:

\begin{itemize}

\item \textit{\textbf{Majorana modes:}} The coupling of Higgs modes to chiral particles can induce mass generation, enabling quantum sensing and switching of chiral Majorana modes. THz-2DCS is uniquely suited to probe these phenomena due to its exceptional sensitivity to Higgs collective modes~\cite{Luo2023,mootz2023multidimensional}. Particularly, THz-2DCM could further enable all-optical quantum sensing and coherent control of Majorana modes with ultrafast timescale, THz frequency precision, and nanometer spatial resolution.

\item \textit{\textbf{Leggett echo:}} The successful demonstration of THz echo spectroscopy in superconductors has revealed unique distinctions arising from the differing nature of Higgs collective modes and Josephson plasmons~\cite{Liu2024,Huang2023}. Building on this foundation, it would be particularly exciting to demonstrate Leggett echo spectroscopy in multiband superconductors. 
Such experiments could uncover quantum pathways associated with Higgs–Leggett and quasiparticle couplings, the broadening and the anharmonicity of Leggett modes, and unambiguous signatures of interband phase coherence. 

\item \textit{\textbf{Quantum magnets:}} While nonlinear magnonic responses have been extensively studied theoretically as probes of fractionalization, decoherence, and wavefunction properties in quantum spin liquids~\cite{Wan2019,Choi2020,Nandkishore2021,negahdariNonlinearResponseKitaev2023,qiang2023probing,Potts2024,zhang2024disentanglingspinexcitationcontinua}, low-dimensional magnets~\cite{liPhotonEchoLensing2021,hartExtractingSpinonSelfenergies2023,gaoTwodimensionalCoherentSpectrum2023,simMicroscopicDetailsTwodimensional2023,liPhotonEchoFractional2023,pottsExploitingPolarizationDependence2023,Watanabe2025,brenig2025twodimensionalnonlinearopticalresponse,Zhang2023DM,srivastava2025theorynonlinearspectroscopyquantum}, and random magnets~\cite{parameswaranAsymptoticallyExactTheory2020}, experimental demonstrations using THz-2DCS remain scarce. The emerging magnets, such as multiferroics \cite{Takahashi2012,Fiebig2016,Li2019,Gao2024} and low-dimensional magnets \cite{Huang2018, Burch2018, Gibertini2019,Gong2019} also remain largely unexplored via THz-2DCS. 
Progress in these directions with Magneto-THz 2DCS could drive transformative advances in spintronics and ultrafast cohrent magnetism \cite{femtomag}.

\item \textit{\textbf{Landau-Dirac fermions:}} Magneto-THz 2DCS provides a unique opportunity to demonstrate and understand extreme nonlinear Landau level mixing in three-dimensional Dirac semimetals~\cite{Jeon2014}. These effects arise from strong electronic nonlinearities and the topological properties of conical electronic states, revealing novel nonlinear phenomena beyond those observed in two-dimensional electron gases in graphene and conventional semiconductors~\cite{Maag2016,Otto2017,Mornhinweg2021}.

\item \textit{\textbf{Axion collective modes:}} Recent advances in emergent low-energy axion electrodynamics within magnetic or strongly correlated topological materials offer a promising pathway for uncovering axion collective modes~\cite{Nenno2020,Qiu2025,Weyl_CDW}. 
These developments hold substantial potential to revolutionize energy and information technologies through THz nonlinear light-matter interactions. Of particular interest is the prediction of dynamic axion insulators, where the interplay between topology and strong interactions gives rise to THz electromagnetic responses and magnetic collective modes \cite{Weyl_CDW}. Magneto-THz 2DCS detection of axion collective modes would enable direct measurement of their quantum excitation pathways and broadening mechanisms.

\item \textit{\textbf{Dynamic Weyl nodes:}} The concept of symmetry control through phononic switches~\cite{luo,vasw2020} can be integrated into topological materials to achieve light-induced Dirac nodes, Weyl nodes, and surface Fermi arcs. THz-2DCS offers compelling opportunities to resolve quantum pathways of phonon excitations and their nonlinear coupling during dynamical topological phase transitions. Extending these studies to cryogenic THz-2DCM enables real-space visualization of hyperbolic and nonreciprocal quantum transport phenomena, including those arising from the light-induced Fermi-arc surface states.

\end{itemize}

In summary, these emerging directions serve as essential bridges across diverse disciplines, linking communities in quantum materials, coherent control, nonlinear and nano-photonics, nonequilibrium physics, spintronics, magnonics, superconducting technologies, topological devices, and quantum information science. Moreover, they inspire visionary experiments designed to transcend traditional boundaries in quantum materials research, pushing toward unprecedented precision in the space–time–frequency domain with near-attojoule energy dissipation. These advances lay the foundation for transformative breakthroughs and highly coherent quantum technologies in the coming years.

\endgroup

\section*{Acknowledgments}
The study was supported by the U.S. Department of Energy,
Office of Basic Energy Science, Division of Materials Sciences and Engineering (Ames National
Laboratory is operated for the U.S. Department of Energy by Iowa State University under Contract
No. DE-AC02-07CH11358).

\section*{Author contributions}
All authors contributed equally to this manuscript.

\section*{Competing interests}
The authors declare no competing interests.


\bibliography{ref}

\begin{thebibliography}{100}
\expandafter\ifx\csname url\endcsname\relax
  \def\url#1{\burl{#1}}\fi
\expandafter\ifx\csname urlprefix\endcsname\relax\def\urlprefix{URL }\fi
\providecommand{\bibinfo}[2]{#2}
\providecommand{\eprint}[2][]{\url{#2}}
\providecommand{\doi}[1]{\url{https://doi.org/#1}}
\bibcommenthead

\bibitem{Basov2017}
\bibinfo{author}{Basov, D.~N.}, \bibinfo{author}{Averitt, R.~D.} \&
  \bibinfo{author}{Hsieh, D.}
\newblock \bibinfo{title}{Towards properties on demand in quantum materials}.
\newblock \emph{\bibinfo{journal}{Nat. Mater.}} \textbf{\bibinfo{volume}{16}},
  \bibinfo{pages}{1077--1088} (\bibinfo{year}{2017}).

\bibitem{Awschalom2018}
\bibinfo{author}{Awschalom, D.~D.}, \bibinfo{author}{Hanson, R.},
  \bibinfo{author}{Wrachtrup, J.} \& \bibinfo{author}{Zhou, B.~B.}
\newblock \bibinfo{title}{Quantum technologies with optically interfaced
  solid-state spins}.
\newblock \emph{\bibinfo{journal}{Nat. Photonics}}
  \textbf{\bibinfo{volume}{12}}, \bibinfo{pages}{516--527}
  (\bibinfo{year}{2018}).

\bibitem{Georgescu2014}
\bibinfo{author}{Georgescu, I.~M.}, \bibinfo{author}{Ashhab, S.} \&
  \bibinfo{author}{Nori, F.}
\newblock \bibinfo{title}{Quantum simulation}.
\newblock \emph{\bibinfo{journal}{Rev. Mod. Phys.}}
  \textbf{\bibinfo{volume}{86}}, \bibinfo{pages}{153} (\bibinfo{year}{2014}).

\bibitem{Preskill2018}
\bibinfo{author}{Preskill, J.}
\newblock \bibinfo{title}{Quantum computing in the {NISQ} era and beyond}.
\newblock \emph{\bibinfo{journal}{Quantum}} \textbf{\bibinfo{volume}{2}},
  \bibinfo{pages}{79} (\bibinfo{year}{2018}).

\bibitem{Gross2017}
\bibinfo{author}{Gross, C.} \& \bibinfo{author}{Bloch, I.}
\newblock \bibinfo{title}{Quantum simulations with ultracold atoms in optical
  lattices}.
\newblock \emph{\bibinfo{journal}{Science}} \textbf{\bibinfo{volume}{357}},
  \bibinfo{pages}{995--1001} (\bibinfo{year}{2017}).

\bibitem{Foss-Feig2025}
\bibinfo{author}{Foss-Feig, M.}, \bibinfo{author}{Pagano, G.},
  \bibinfo{author}{Potter, A.~C.} \& \bibinfo{author}{Yao, N.~Y.}
\newblock \bibinfo{title}{Progress in trapped-ion quantum simulation}.
\newblock \emph{\bibinfo{journal}{Annu. Rev. Condens. Matter Phys.}}
  \textbf{\bibinfo{volume}{16}}, \bibinfo{pages}{145--172}
  (\bibinfo{year}{2025}).

\bibitem{Warner2025}
\bibinfo{author}{Warner, H.~K.} \emph{et~al.}
\newblock \bibinfo{title}{Coherent control of a superconducting qubit using
  light}.
\newblock \emph{\bibinfo{journal}{Nat. Phys.}}  (\bibinfo{year}{2025}).

\bibitem{Elsaesser2Dbook}
\bibinfo{author}{Elsaesser, T.}, \bibinfo{author}{Reimann, K.} \&
  \bibinfo{author}{Woerner, M.}
\newblock \emph{\bibinfo{title}{Concepts and {Applications} of {Nonlinear
  Terahertz Spectroscopy}}}  (\bibinfo{publisher}{Morgan \& Claypool
  Publishers}, \bibinfo{year}{2019}).

\bibitem{Nicoletti:16}
\bibinfo{author}{Nicoletti, D.} \& \bibinfo{author}{Cavalleri, A.}
\newblock \bibinfo{title}{Nonlinear light--matter interaction at terahertz
  frequencies}.
\newblock \emph{\bibinfo{journal}{Adv. Opt. Photon.}}
  \textbf{\bibinfo{volume}{8}}, \bibinfo{pages}{401--464}
  (\bibinfo{year}{2016}).

\bibitem{bristow2009versatile}
\bibinfo{author}{Bristow, A.} \emph{et~al.}
\newblock \bibinfo{title}{A versatile ultrastable platform for optical
  multidimensional fourier-transform spectroscopy}.
\newblock \emph{\bibinfo{journal}{Rev. Sci. Instrum.}}
  \textbf{\bibinfo{volume}{80}}, \bibinfo{pages}{073108}
  (\bibinfo{year}{2009}).

\bibitem{Cundiff2013}
\bibinfo{author}{Cundiff, S.~T.} \& \bibinfo{author}{Mukamel, S.}
\newblock \bibinfo{title}{{Optical multidimensional coherent spectroscopy}}.
\newblock \emph{\bibinfo{journal}{Physics Today}}
  \textbf{\bibinfo{volume}{66}}, \bibinfo{pages}{44--49}
  (\bibinfo{year}{2013}).

\bibitem{li2006many}
\bibinfo{author}{Li, X.}, \bibinfo{author}{Zhang, T.}, \bibinfo{author}{Borca,
  C.~N.} \& \bibinfo{author}{Cundiff, S.~T.}
\newblock \bibinfo{title}{Many-body interactions in semiconductors probed by
  optical two-dimensional {Fourier} transform spectroscopy}.
\newblock \emph{\bibinfo{journal}{Phys. Rev. Lett.}}
  \textbf{\bibinfo{volume}{96}}, \bibinfo{pages}{057406}
  (\bibinfo{year}{2006}).

\bibitem{yang2007two}
\bibinfo{author}{Yang, L.}, \bibinfo{author}{Schweigert, I.~V.},
  \bibinfo{author}{Cundiff, S.~T.} \& \bibinfo{author}{Mukamel, S.}
\newblock \bibinfo{title}{Two-dimensional optical spectroscopy of excitons in
  semiconductor quantum wells: Liouville-space pathway analysis}.
\newblock \emph{\bibinfo{journal}{Phys. Rev. B}} \textbf{\bibinfo{volume}{75}},
  \bibinfo{pages}{125302} (\bibinfo{year}{2007}).

\bibitem{li2013unraveling}
\bibinfo{author}{Li, H.}, \bibinfo{author}{Bristow, A.~D.},
  \bibinfo{author}{Siemens, M.~E.}, \bibinfo{author}{Moody, G.} \&
  \bibinfo{author}{Cundiff, S.~T.}
\newblock \bibinfo{title}{Unraveling quantum pathways using optical {3D}
  fourier-transform spectroscopy}.
\newblock \emph{\bibinfo{journal}{Nat. Commun.}} \textbf{\bibinfo{volume}{4}},
  \bibinfo{pages}{1--9} (\bibinfo{year}{2013}).

\bibitem{cowan2005ultrafast}
\bibinfo{author}{Cowan, M.} \emph{et~al.}
\newblock \bibinfo{title}{Ultrafast memory loss and energy redistribution in
  the hydrogen bond network of liquid {H$_2$O}}.
\newblock \emph{\bibinfo{journal}{Nature}} \textbf{\bibinfo{volume}{434}},
  \bibinfo{pages}{199--202} (\bibinfo{year}{2005}).

\bibitem{dahms2017large}
\bibinfo{author}{Dahms, F.}, \bibinfo{author}{Fingerhut, B.~P.},
  \bibinfo{author}{Nibbering, E.~T.}, \bibinfo{author}{Pines, E.} \&
  \bibinfo{author}{Elsaesser, T.}
\newblock \bibinfo{title}{Large-amplitude transfer motion of hydrated excess
  protons mapped by ultrafast {2D IR} spectroscopy}.
\newblock \emph{\bibinfo{journal}{Science}} \textbf{\bibinfo{volume}{357}},
  \bibinfo{pages}{491--495} (\bibinfo{year}{2017}).

\bibitem{Yang2023}
\bibinfo{author}{Yang, C.~J.}, \bibinfo{author}{Li, J.},
  \bibinfo{author}{Fiebig, M.} \& \bibinfo{author}{Pal, S.}
\newblock \bibinfo{title}{Terahertz control of many-body dynamics in quantum
  materials}.
\newblock \emph{\bibinfo{journal}{Nat. Rev. Mater.}}
  \textbf{\bibinfo{volume}{8}}, \bibinfo{pages}{518--532}
  (\bibinfo{year}{2023}).

\bibitem{yang2018}
\bibinfo{author}{Yang, X.} \emph{et~al.}
\newblock \bibinfo{title}{Terahertz-light quantum tuning of a metastable
  emergent phase hidden by superconductivity}.
\newblock \emph{\bibinfo{journal}{Nat. Mater.}} \textbf{\bibinfo{volume}{17}},
  \bibinfo{pages}{586} (\bibinfo{year}{2018}).

\bibitem{yang2019lightwave}
\bibinfo{author}{Yang, X.} \emph{et~al.}
\newblock \bibinfo{title}{Lightwave-driven superconductivity and forbidden
  quantum beats by terahertz symmetry breaking}.
\newblock \emph{\bibinfo{journal}{Nat. Photonics}}
  \textbf{\bibinfo{volume}{13}}, \bibinfo{pages}{707--713}
  (\bibinfo{year}{2019}).

\bibitem{Luo2023}
\bibinfo{author}{Luo, L.} \emph{et~al.}
\newblock \bibinfo{title}{Quantum coherence tomography of light-controlled
  superconductivity}.
\newblock \emph{\bibinfo{journal}{Nat. Phys.}} \textbf{\bibinfo{volume}{19}},
  \bibinfo{pages}{201--209} (\bibinfo{year}{2023}).

\bibitem{Lu2019}
\bibinfo{author}{Lu, J.} \emph{et~al.}
\newblock \emph{\bibinfo{title}{Two-Dimensional Spectroscopy at Terahertz
  Frequencies}}, \bibinfo{pages}{275--320} (\bibinfo{publisher}{Springer
  International Publishing}, \bibinfo{year}{2019}).

\bibitem{Liu2025}
\bibinfo{author}{Liu, A.}
\newblock \bibinfo{title}{Multidimensional terahertz probes of quantum
  materials}.
\newblock \emph{\bibinfo{journal}{npj Quantum Mater.}}
  \textbf{\bibinfo{volume}{10}}, \bibinfo{pages}{18} (\bibinfo{year}{2025}).

\bibitem{Huang2023}
\bibinfo{author}{Huang, C.} \emph{et~al.}
\newblock \bibinfo{title}{Discovery of an {Unconventional Quantum Echo} by
  {Interference} of {Higgs Coherence}}.
\newblock \emph{\bibinfo{journal}{Sci. Adv.}} \textbf{\bibinfo{volume}{11}},
  \bibinfo{pages}{eads8740} (\bibinfo{year}{2025}).

\bibitem{vaswani2019discovery}
\bibinfo{author}{Vaswani, C.} \emph{et~al.}
\newblock \bibinfo{title}{Terahertz second-harmonic generation from lightwave
  acceleration of symmetry-breaking nonlinear supercurrents}.
\newblock \emph{\bibinfo{journal}{Phys. Rev. Lett.}}
  \textbf{\bibinfo{volume}{124}}, \bibinfo{pages}{207003}
  (\bibinfo{year}{2020}).

\bibitem{Mootz2020}
\bibinfo{author}{Mootz, M.}, \bibinfo{author}{Wang, J.} \&
  \bibinfo{author}{Perakis, I.~E.}
\newblock \bibinfo{title}{Lightwave terahertz quantum manipulation of
  nonequilibrium superconductor phases and their collective modes}.
\newblock \emph{\bibinfo{journal}{Phys. Rev. B}}
  \textbf{\bibinfo{volume}{102}}, \bibinfo{pages}{054517}
  (\bibinfo{year}{2020}).

\bibitem{luo}
\bibinfo{author}{Luo, L.} \emph{et~al.}
\newblock \bibinfo{title}{A light-induced phononic symmetry switch and giant
  dissipationless topological photocurrent in {ZrTe$_5$}}.
\newblock \emph{\bibinfo{journal}{Nat. Mater.}} \textbf{\bibinfo{volume}{20}},
  \bibinfo{pages}{329--334} (\bibinfo{year}{2021}).

\bibitem{vasw2020}
\bibinfo{author}{Vaswani, C.} \emph{et~al.}
\newblock \bibinfo{title}{Light-driven {Raman} coherence as a nonthermal route
  to ultrafast topology switching in a {Dirac} semimetal}.
\newblock \emph{\bibinfo{journal}{Phys. Rev. X}} \textbf{\bibinfo{volume}{10}},
  \bibinfo{pages}{021013} (\bibinfo{year}{2020}).

\bibitem{Yang2020}
\bibinfo{author}{Yang, X.} \emph{et~al.}
\newblock \bibinfo{title}{Light control of surface--bulk coupling by terahertz
  vibrational coherence in a topological insulator}.
\newblock \emph{\bibinfo{journal}{npj Quantum Mater.}}
  \textbf{\bibinfo{volume}{5}}, \bibinfo{pages}{13} (\bibinfo{year}{2020}).

\bibitem{Huang2024}
\bibinfo{author}{Huang, C.} \emph{et~al.}
\newblock \bibinfo{title}{Extreme terahertz magnon multiplication induced by
  resonant magnetic pulse pairs}.
\newblock \emph{\bibinfo{journal}{Nat. Commun.}} \textbf{\bibinfo{volume}{15}},
  \bibinfo{pages}{3214} (\bibinfo{year}{2024}).

\bibitem{mootz2023multidimensional}
\bibinfo{author}{Mootz, M.}, \bibinfo{author}{Luo, L.}, \bibinfo{author}{Huang,
  C.}, \bibinfo{author}{Wang, J.} \& \bibinfo{author}{Perakis, l.~E.}
\newblock \bibinfo{title}{Multidimensional coherent spectroscopy of
  light-driven states and their collective modes in multiband superconductors}.
\newblock \emph{\bibinfo{journal}{Phys. Rev. B}}
  \textbf{\bibinfo{volume}{109}}, \bibinfo{pages}{014515}
  (\bibinfo{year}{2024}).

\bibitem{Kuehn2009}
\bibinfo{author}{Kuehn, W.}, \bibinfo{author}{Reimann, K.},
  \bibinfo{author}{Woerner, M.} \& \bibinfo{author}{Elsaesser, T.}
\newblock \bibinfo{title}{Phase-resolved two-dimensional spectroscopy based on
  collinear $n$-wave mixing in the ultrafast time domain}.
\newblock \emph{\bibinfo{journal}{J. Chem. Phys.}}
  \textbf{\bibinfo{volume}{130}}, \bibinfo{pages}{164503}
  (\bibinfo{year}{2009}).

\bibitem{Kuehn2011QW1}
\bibinfo{author}{Kuehn, W.}, \bibinfo{author}{Reimann, K.},
  \bibinfo{author}{Woerner, M.}, \bibinfo{author}{Elsaesser, T.} \&
  \bibinfo{author}{Hey, R.}
\newblock \bibinfo{title}{Two-{Dimensional Terahertz Correlation Spectra} of
  {Electronic Excitations} in {Semiconductor Quantum Wells}}.
\newblock \emph{\bibinfo{journal}{J. Phys. Chem. B}}
  \textbf{\bibinfo{volume}{115}}, \bibinfo{pages}{5448--5455}
  (\bibinfo{year}{2011}).

\bibitem{Junginger2012}
\bibinfo{author}{Junginger, F.} \emph{et~al.}
\newblock \bibinfo{title}{Nonperturbative {Interband Response} of a {Bulk InSb
  Semiconductor Driven Off Resonantly} by {Terahertz Electromagnetic Few-Cycle
  Pulses}}.
\newblock \emph{\bibinfo{journal}{Phys. Rev. Lett.}}
  \textbf{\bibinfo{volume}{109}}, \bibinfo{pages}{147403}
  (\bibinfo{year}{2012}).

\bibitem{Woerner2013}
\bibinfo{author}{Woerner, M.}, \bibinfo{author}{Kuehn, W.},
  \bibinfo{author}{Bowlan, P.}, \bibinfo{author}{Reimann, K.} \&
  \bibinfo{author}{Elsaesser, T.}
\newblock \bibinfo{title}{Ultrafast two-dimensional terahertz spectroscopy of
  elementary excitations in solids}.
\newblock \emph{\bibinfo{journal}{New J. Phys.}} \textbf{\bibinfo{volume}{15}},
  \bibinfo{pages}{025039} (\bibinfo{year}{2013}).

\bibitem{Woerner2019}
\bibinfo{author}{Woerner, M.} \emph{et~al.}
\newblock \bibinfo{title}{Terahertz {Driven Amplification} of {Coherent Optical
  Phonons} in {GaAs Coupled} to a {Metasurface}}.
\newblock \emph{\bibinfo{journal}{Phys. Rev. Lett.}}
  \textbf{\bibinfo{volume}{122}}, \bibinfo{pages}{107402}
  (\bibinfo{year}{2019}).

\bibitem{Ghalgaoui2020GaAs}
\bibinfo{author}{Ghalgaoui, A.} \emph{et~al.}
\newblock \bibinfo{title}{Frequency {Upshift} of the {Transverse Optical Phonon
  Resonance} in {GaAs} by {Femtosecond Electron-Hole Excitation}}.
\newblock \emph{\bibinfo{journal}{Phys. Rev. Lett.}}
  \textbf{\bibinfo{volume}{125}}, \bibinfo{pages}{027401}
  (\bibinfo{year}{2020}).

\bibitem{Woerner2021}
\bibinfo{author}{Woerner, M.}, \bibinfo{author}{Ghalgaoui, A.},
  \bibinfo{author}{Reimann, K.} \& \bibinfo{author}{Elsaesser, T.}
\newblock \bibinfo{title}{Two-color two-dimensional terahertz spectroscopy: A
  new approach for exploring even-order nonlinearities in the nonperturbative
  regime}.
\newblock \emph{\bibinfo{journal}{J. Chem. Phys.}}
  \textbf{\bibinfo{volume}{154}}, \bibinfo{pages}{154203}
  (\bibinfo{year}{2021}).

\bibitem{Kuehn2011QW2}
\bibinfo{author}{Kuehn, W.} \emph{et~al.}
\newblock \bibinfo{title}{Strong {Correlation} of {Electronic} and {Lattice
  Excitations} in {GaAs}/{AlGaAs} {Semiconductor Quantum Wells Revealed} by
  {Two-Dimensional Terahertz Spectroscopy}}.
\newblock \emph{\bibinfo{journal}{Phys. Rev. Lett.}}
  \textbf{\bibinfo{volume}{107}}, \bibinfo{pages}{067401}
  (\bibinfo{year}{2011}).

\bibitem{Bowlan2014}
\bibinfo{author}{Bowlan, P.}, \bibinfo{author}{{Martinez-Moreno}, E.},
  \bibinfo{author}{Reimann, K.}, \bibinfo{author}{Elsaesser, T.} \&
  \bibinfo{author}{Woerner, M.}
\newblock \bibinfo{title}{{Ultrafast Terahertz Response of Multilayer Graphene
  in the Nonperturbative Regime}}.
\newblock \emph{\bibinfo{journal}{Phys. Rev. B}} \textbf{\bibinfo{volume}{89}},
  \bibinfo{pages}{041408} (\bibinfo{year}{2014}).

\bibitem{Somma2016InSb}
\bibinfo{author}{Somma, C.}, \bibinfo{author}{Folpini, G.},
  \bibinfo{author}{Reimann, K.}, \bibinfo{author}{Woerner, M.} \&
  \bibinfo{author}{Elsaesser, T.}
\newblock \bibinfo{title}{Two-{Phonon Quantum Coherences} in {Indium Antimonide
  Studied} by {Nonlinear Two-Dimensional Terahertz Spectroscopy}}.
\newblock \emph{\bibinfo{journal}{Phys. Rev. Lett.}}
  \textbf{\bibinfo{volume}{116}}, \bibinfo{pages}{177401}
  (\bibinfo{year}{2016}).

\bibitem{Somma2016beyond}
\bibinfo{author}{Somma, C.}, \bibinfo{author}{Folpini, G.},
  \bibinfo{author}{Reimann, K.}, \bibinfo{author}{Woerner, M.} \&
  \bibinfo{author}{Elsaesser, T.}
\newblock \bibinfo{title}{Phase-resolved two-dimensional terahertz spectroscopy
  including off-resonant interactions beyond the {\emph{$\chi^{(3)}$}} limit}.
\newblock \emph{\bibinfo{journal}{J. Chem. Phys}}
  \textbf{\bibinfo{volume}{144}}, \bibinfo{pages}{184202}
  (\bibinfo{year}{2016}).

\bibitem{Houver2019}
\bibinfo{author}{Houver, S.}, \bibinfo{author}{Huber, L.},
  \bibinfo{author}{Savoini, M.}, \bibinfo{author}{Abreu, E.} \&
  \bibinfo{author}{Johnson, S.~L.}
\newblock \bibinfo{title}{{2D THz} spectroscopic investigation of ballistic
  conduction-band electron dynamics in {InSb}}.
\newblock \emph{\bibinfo{journal}{Opt. Express}} \textbf{\bibinfo{volume}{27}},
  \bibinfo{pages}{10854} (\bibinfo{year}{2019}).

\bibitem{Markmann2020}
\bibinfo{author}{Markmann, S.} \emph{et~al.}
\newblock \bibinfo{title}{Two-dimensional spectroscopy on a {THz} quantum
  cascade structure}.
\newblock \emph{\bibinfo{journal}{Nanophotonics}}
  \textbf{\bibinfo{volume}{10}}, \bibinfo{pages}{171--180}
  (\bibinfo{year}{2020}).

\bibitem{Raab2019}
\bibinfo{author}{Raab, J.} \emph{et~al.}
\newblock \bibinfo{title}{Ultrafast two-dimensional field spectroscopy of
  terahertz intersubband saturable absorbers}.
\newblock \emph{\bibinfo{journal}{Opt. Express}} \textbf{\bibinfo{volume}{27}},
  \bibinfo{pages}{2248} (\bibinfo{year}{2019}).

\bibitem{Raab2020}
\bibinfo{author}{Raab, J.} \emph{et~al.}
\newblock \bibinfo{title}{Ultrafast terahertz saturable absorbers using
  tailored intersubband polaritons}.
\newblock \emph{\bibinfo{journal}{Nat. Commun.}} \textbf{\bibinfo{volume}{11}},
  \bibinfo{pages}{4290} (\bibinfo{year}{2020}).

\bibitem{Tarekegne2020}
\bibinfo{author}{Tarekegne, A.~T.} \emph{et~al.}
\newblock \bibinfo{title}{Subcycle {Nonlinear Response} of {Doped} 4{$H$}
  {Silicon Carbide Revealed} by {Two-Dimensional Terahertz Spectroscopy}}.
\newblock \emph{\bibinfo{journal}{ACS Photonics}} \textbf{\bibinfo{volume}{7}},
  \bibinfo{pages}{221--231} (\bibinfo{year}{2020}).

\bibitem{Mahmood2021}
\bibinfo{author}{Mahmood, F.}, \bibinfo{author}{Chaudhuri, D.},
  \bibinfo{author}{Gopalakrishnan, S.}, \bibinfo{author}{Nandkishore, R.} \&
  \bibinfo{author}{Armitage, N.~P.}
\newblock \bibinfo{title}{Observation of a marginal {Fermi} glass}.
\newblock \emph{\bibinfo{journal}{Nat. Phys.}} \textbf{\bibinfo{volume}{17}},
  \bibinfo{pages}{627--631} (\bibinfo{year}{2021}).

\bibitem{Riepl2021}
\bibinfo{author}{Riepl, J.} \emph{et~al.}
\newblock \bibinfo{title}{Field-resolved high-order sub-cycle nonlinearities in
  a terahertz semiconductor laser}.
\newblock \emph{\bibinfo{journal}{Light Sci. Appl.}}
  \textbf{\bibinfo{volume}{10}}, \bibinfo{pages}{246} (\bibinfo{year}{2021}).

\bibitem{Gill2024}
\bibinfo{author}{Gill, T.~B.} \emph{et~al.}
\newblock \bibinfo{title}{{2D Time-Domain Spectroscopy} for {Determination} of
  {Energy} and {Momentum Relaxation Rates} of {Hydrogen-Like Donor States} in
  {Germanium}}.
\newblock \emph{\bibinfo{journal}{ACS Photonics}}
  \textbf{\bibinfo{volume}{11}}, \bibinfo{pages}{1447--1455}
  (\bibinfo{year}{2024}).

\bibitem{Elsaesser2015}
\bibinfo{author}{Elsaesser, T.}, \bibinfo{author}{Reimann, K.} \&
  \bibinfo{author}{Woerner, M.}
\newblock \bibinfo{title}{{Focus: Phase-resolved Nonlinear Terahertz
  Spectroscopy--From Charge Dynamics in Solids to Molecular Excitations in
  Liquids}}.
\newblock \emph{\bibinfo{journal}{J. Chem. Phys}}
  \textbf{\bibinfo{volume}{142}}, \bibinfo{pages}{212301}
  (\bibinfo{year}{2015}).

\bibitem{Maag2016}
\bibinfo{author}{Maag, T.} \emph{et~al.}
\newblock \bibinfo{title}{Coherent cyclotron motion beyond {Kohn}'s theorem}.
\newblock \emph{\bibinfo{journal}{Nat. Phys.}} \textbf{\bibinfo{volume}{12}},
  \bibinfo{pages}{119--123} (\bibinfo{year}{2016}).

\bibitem{Mornhinweg2021}
\bibinfo{author}{Mornhinweg, J.} \emph{et~al.}
\newblock \bibinfo{title}{Tailored {Subcycle Nonlinearities} of {Ultrastrong
  Light-Matter Coupling}}.
\newblock \emph{\bibinfo{journal}{Phys. Rev. Lett.}}
  \textbf{\bibinfo{volume}{126}}, \bibinfo{pages}{177404}
  (\bibinfo{year}{2021}).

\bibitem{Pashkin2013}
\bibinfo{author}{Pashkin, A.}, \bibinfo{author}{Sell, A.},
  \bibinfo{author}{Kampfrath, T.} \& \bibinfo{author}{Huber, R.}
\newblock \bibinfo{title}{Electric and magnetic terahertz nonlinearities
  resolved on the sub-cycle scale}.
\newblock \emph{\bibinfo{journal}{New J. Phys.}} \textbf{\bibinfo{volume}{15}},
  \bibinfo{pages}{065003} (\bibinfo{year}{2013}).

\bibitem{Lu2017}
\bibinfo{author}{Lu, J.} \emph{et~al.}
\newblock \bibinfo{title}{Coherent {Two-Dimensional Terahertz Magnetic
  Resonance Spectroscopy} of {Collective Spin Waves}}.
\newblock \emph{\bibinfo{journal}{Phys. Rev. Lett.}}
  \textbf{\bibinfo{volume}{118}}, \bibinfo{pages}{207204}
  (\bibinfo{year}{2017}).

\bibitem{Dutta2025}
\bibinfo{author}{Dutta, A.} \emph{et~al.}
\newblock \bibinfo{title}{Evidence of {Relativistic Field}-{Derivative Torque}
  in {Nonlinear THz Response} of {Magnetization Dynamics}}.
\newblock \emph{\bibinfo{journal}{Adv. Funct. Mater.}}
  \textbf{\bibinfo{volume}{35}}, \bibinfo{pages}{2414582}
  (\bibinfo{year}{2025}).

\bibitem{Mashkovich2021}
\bibinfo{author}{Mashkovich, E.~A.} \emph{et~al.}
\newblock \bibinfo{title}{Terahertz light--driven coupling of antiferromagnetic
  spins to lattice}.
\newblock \emph{\bibinfo{journal}{Science}} \textbf{\bibinfo{volume}{374}},
  \bibinfo{pages}{1608--1611} (\bibinfo{year}{2021}).

\bibitem{Grishunin2023}
\bibinfo{author}{Grishunin, K.~A.}, \bibinfo{author}{Bilyk, V.~R.},
  \bibinfo{author}{Mishina, E.~D.}, \bibinfo{author}{Kimel, A.~V.} \&
  \bibinfo{author}{Mashkovich, E.~A.}
\newblock \bibinfo{title}{Two-dimensional terahertz spectroscopy as a tool for
  revealing nonlinear interactions in media}.
\newblock \emph{\bibinfo{journal}{Rev. Sci. Instrum.}}
  \textbf{\bibinfo{volume}{94}}, \bibinfo{pages}{073005}
  (\bibinfo{year}{2023}).

\bibitem{Blank2023Spin}
\bibinfo{author}{Blank, T. G.~H.} \emph{et~al.}
\newblock \bibinfo{title}{Empowering {Control} of {Antiferromagnets} by
  {THz-Induced Spin Coherence}}.
\newblock \emph{\bibinfo{journal}{Phys. Rev. Lett.}}
  \textbf{\bibinfo{volume}{131}}, \bibinfo{pages}{096701}
  (\bibinfo{year}{2023}).

\bibitem{Zhang2024Coup}
\bibinfo{author}{Zhang, Z.} \emph{et~al.}
\newblock \bibinfo{title}{Terahertz field-induced nonlinear coupling of two
  magnon modes in an antiferromagnet}.
\newblock \emph{\bibinfo{journal}{Nat. Phys.}} \textbf{\bibinfo{volume}{20}},
  \bibinfo{pages}{801--806} (\bibinfo{year}{2024}).

\bibitem{Zhang2024Down}
\bibinfo{author}{Zhang, Z.} \emph{et~al.}
\newblock \bibinfo{title}{Terahertz stimulated parametric downconversion of a
  magnon mode in an antiferromagnet}.
\newblock \emph{\bibinfo{journal}{Sci. Adv.}} \textbf{\bibinfo{volume}{11}},
  \bibinfo{pages}{eadv3757} (\bibinfo{year}{2025}).

\bibitem{Zhang2024Up}
\bibinfo{author}{Zhang, Z.} \emph{et~al.}
\newblock \bibinfo{title}{Terahertz-field-driven magnon upconversion in an
  antiferromagnet}.
\newblock \emph{\bibinfo{journal}{Nat. Phys.}} \textbf{\bibinfo{volume}{20}},
  \bibinfo{pages}{788--793} (\bibinfo{year}{2024}).

\bibitem{Leenders2024}
\bibinfo{author}{Leenders, R.~A.}, \bibinfo{author}{Afanasiev, D.},
  \bibinfo{author}{Kimel, A.~V.} \& \bibinfo{author}{Mikhaylovskiy, R.~V.}
\newblock \bibinfo{title}{Canted spin order as a platform for ultrafast
  conversion of magnons}.
\newblock \emph{\bibinfo{journal}{Nature}} \textbf{\bibinfo{volume}{630}},
  \bibinfo{pages}{335--339} (\bibinfo{year}{2024}).

\bibitem{Zhang2024Spin}
\bibinfo{author}{Zhang, Z.} \emph{et~al.}
\newblock \bibinfo{title}{Spin switching in {Sm$_{0.7}$Er$_{0.3}$FeO$_3$}
  triggered by terahertz magnetic-field pulses}.
\newblock \emph{\bibinfo{journal}{Nat. Mater.}} \textbf{\bibinfo{volume}{24}},
  \bibinfo{pages}{219--225} (\bibinfo{year}{2024}).

\bibitem{Katsumi2024MgB2}
\bibinfo{author}{Katsumi, K.} \emph{et~al.}
\newblock \bibinfo{title}{Amplitude mode in a multi-gap superconductor
  {MgB}$_2$ investigated by terahertz two-dimensional coherent spectroscopy}
  (\bibinfo{year}{2024}).
\newblock \eprint{arXiv:2411.10852}.

\bibitem{Kim2024}
\bibinfo{author}{Kim, M.~J.} \emph{et~al.}
\newblock \bibinfo{title}{Tracing the dynamics of superconducting order via
  transient terahertz third-harmonic generation}.
\newblock \emph{\bibinfo{journal}{Sci. Adv.}} \textbf{\bibinfo{volume}{10}},
  \bibinfo{pages}{eadi7598} (\bibinfo{year}{2024}).

\bibitem{Katsumi2024NbN}
\bibinfo{author}{Katsumi, K.} \emph{et~al.}
\newblock \bibinfo{title}{Revealing {Novel Aspects} of {Light-Matter Coupling}
  by {Terahertz Two-Dimensional Coherent Spectroscopy}: {The Case} of the
  {Amplitude Mode} in {Superconductors}}.
\newblock \emph{\bibinfo{journal}{Phys. Rev. Lett.}}
  \textbf{\bibinfo{volume}{132}}, \bibinfo{pages}{256903}
  (\bibinfo{year}{2024}).

\bibitem{Liu2024}
\bibinfo{author}{Liu, A.} \emph{et~al.}
\newblock \bibinfo{title}{Probing inhomogeneous cuprate superconductivity by
  terahertz {Josephson} echo spectroscopy}.
\newblock \emph{\bibinfo{journal}{Nat. Phys.}} \textbf{\bibinfo{volume}{20}},
  \bibinfo{pages}{1751--1756} (\bibinfo{year}{2024}).

\bibitem{Salvador2024}
\bibinfo{author}{G{\'o}mez~Salvador, A.} \emph{et~al.}
\newblock \bibinfo{title}{Principles of two-dimensional terahertz spectroscopy
  of collective excitations: The case of {Josephson} plasmons in layered
  superconductors}.
\newblock \emph{\bibinfo{journal}{Phys. Rev. B}}
  \textbf{\bibinfo{volume}{110}}, \bibinfo{pages}{094514}
  (\bibinfo{year}{2024}).

\bibitem{chaudhuri2025}
\bibinfo{author}{Chaudhuri, D.} \emph{et~al.}
\newblock \bibinfo{title}{Planckian dissipation, anomalous high temperature
  {THz} non-linear response and energy relaxation in the strange metal state of
  the cuprate superconductors} (\bibinfo{year}{2025}).
\newblock \eprint{arXiv:2503.15646}.

\bibitem{Cheng2025}
\bibinfo{author}{Cheng, B.} \emph{et~al.}
\newblock \bibinfo{title}{Observation of cupratelike nonlinear terahertz
  responses in superconducting infinite-layer nickelates via two-dimensional
  coherent spectroscopy}.
\newblock \emph{\bibinfo{journal}{Phys. Rev. B}}
  \textbf{\bibinfo{volume}{111}}, \bibinfo{pages}{014519}
  (\bibinfo{year}{2025}).

\bibitem{Blank2023Phonon}
\bibinfo{author}{Blank, T. G.~H.} \emph{et~al.}
\newblock \bibinfo{title}{Two-{Dimensional Terahertz Spectroscopy} of
  {Nonlinear Phononics} in the {Topological Insulator MnBi}$_2${Te}$_4$}.
\newblock \emph{\bibinfo{journal}{Phys. Rev. Lett.}}
  \textbf{\bibinfo{volume}{131}}, \bibinfo{pages}{026902}
  (\bibinfo{year}{2023}).

\bibitem{Bhandia2024}
\bibinfo{author}{Bhandia, R.} \emph{et~al.}
\newblock \bibinfo{title}{Anomalous electronic energy relaxation and soft
  phonons in the {Dirac} semimetal {Cd}$_3${As}$_2$}.
\newblock \emph{\bibinfo{journal}{Phys. Rev. B}}
  \textbf{\bibinfo{volume}{110}}, \bibinfo{pages}{075131}
  (\bibinfo{year}{2024}).

\bibitem{Barbalas2025}
\bibinfo{author}{Barbalas, D.} \emph{et~al.}
\newblock \bibinfo{title}{Energy {Relaxation} and {Dynamics} in the {Correlated
  Metal Sr}$_2${RuO}$_4$ via {Terahertz Two-Dimensional Coherent
  Spectroscopy}}.
\newblock \emph{\bibinfo{journal}{Phys. Rev. Lett.}}
  \textbf{\bibinfo{volume}{134}}, \bibinfo{pages}{036501}
  (\bibinfo{year}{2025}).

\bibitem{Somma2014}
\bibinfo{author}{Somma, C.}, \bibinfo{author}{Reimann, K.},
  \bibinfo{author}{Flytzanis, C.}, \bibinfo{author}{Elsaesser, T.} \&
  \bibinfo{author}{Woerner, M.}
\newblock \bibinfo{title}{High-{Field Terahertz Bulk Photovoltaic Effect} in
  {Lithium Niobate}}.
\newblock \emph{\bibinfo{journal}{Phys. Rev. Lett.}}
  \textbf{\bibinfo{volume}{112}}, \bibinfo{pages}{146602}
  (\bibinfo{year}{2014}).

\bibitem{Pal2021}
\bibinfo{author}{Pal, S.} \emph{et~al.}
\newblock \bibinfo{title}{Origin of {Terahertz Soft-Mode Nonlinearities} in
  {Ferroelectric Perovskites}}.
\newblock \emph{\bibinfo{journal}{Phys. Rev. X}} \textbf{\bibinfo{volume}{11}},
  \bibinfo{pages}{021023} (\bibinfo{year}{2021}).

\bibitem{Lin2022}
\bibinfo{author}{Lin, H.~W.}, \bibinfo{author}{Mead, G.} \&
  \bibinfo{author}{Blake, G.~A.}
\newblock \bibinfo{title}{Mapping {LiNbO}$_3$ {Phonon-Polariton Nonlinearities}
  with {2D THz-THz-Raman Spectroscopy}}.
\newblock \emph{\bibinfo{journal}{Phys. Rev. Lett.}}
  \textbf{\bibinfo{volume}{129}}, \bibinfo{pages}{207401}
  (\bibinfo{year}{2022}).

\bibitem{Reimann2021}
\bibinfo{author}{Reimann, K.}, \bibinfo{author}{Woerner, M.} \&
  \bibinfo{author}{Elsaesser, T.}
\newblock \bibinfo{title}{{Two-Dimensional Terahertz Spectroscopy of
  Condensed-Phase Molecular Systems}}.
\newblock \emph{\bibinfo{journal}{J. Chem. Phys}}
  \textbf{\bibinfo{volume}{154}}, \bibinfo{pages}{120901}
  (\bibinfo{year}{2021}).

\bibitem{Folpini2017}
\bibinfo{author}{Folpini, G.} \emph{et~al.}
\newblock \bibinfo{title}{Strong {Local-Field Enhancement} of the {Nonlinear
  Soft-Mode Response} in a {Molecular Crystal}}.
\newblock \emph{\bibinfo{journal}{Phys. Rev. Lett.}}
  \textbf{\bibinfo{volume}{119}}, \bibinfo{pages}{097404}
  (\bibinfo{year}{2017}).

\bibitem{Lu2016}
\bibinfo{author}{Lu, J.} \emph{et~al.}
\newblock \bibinfo{title}{{Nonlinear Two-Dimensional Terahertz Photon Echo and
  Rotational Spectroscopy in the Gas Phase}}.
\newblock \emph{\bibinfo{journal}{Proc. Natl. Acad. Sci. U.S.A.}}
  \textbf{\bibinfo{volume}{113}}, \bibinfo{pages}{11800--11805}
  (\bibinfo{year}{2016}).

\bibitem{Zhang2021}
\bibinfo{author}{Zhang, Y.} \emph{et~al.}
\newblock \bibinfo{title}{{Nonlinear Rotational Spectroscopy Reveals Many-Body
  Interactions in Water Molecules}}.
\newblock \emph{\bibinfo{journal}{Proc. Natl. Acad. Sci. U.S.A.}}
  \textbf{\bibinfo{volume}{118}}, \bibinfo{pages}{e2020941118}
  (\bibinfo{year}{2021}).

\bibitem{Allodi2015}
\bibinfo{author}{Allodi, M.~A.}, \bibinfo{author}{Finneran, I.~A.} \&
  \bibinfo{author}{Blake, G.~A.}
\newblock \bibinfo{title}{{Nonlinear Terahertz Coherent Excitation of
  Vibrational Modes of Liquids}}.
\newblock \emph{\bibinfo{journal}{J. Chem. Phys.}}
  \textbf{\bibinfo{volume}{143}}, \bibinfo{pages}{234204}
  (\bibinfo{year}{2015}).

\bibitem{Finneran2016}
\bibinfo{author}{Finneran, I.~A.}, \bibinfo{author}{Welsch, R.},
  \bibinfo{author}{Allodi, M.~A.}, \bibinfo{author}{Miller, T.~F.} \&
  \bibinfo{author}{Blake, G.~A.}
\newblock \bibinfo{title}{{Coherent Two-Dimensional Terahertz-Terahertz-Raman
  Spectroscopy}}.
\newblock \emph{\bibinfo{journal}{Proc. Natl. Acad. Sci. U.S.A.}}
  \textbf{\bibinfo{volume}{113}}, \bibinfo{pages}{6857--6861}
  (\bibinfo{year}{2016}).

\bibitem{Finneran2017}
\bibinfo{author}{Finneran, I.~A.}, \bibinfo{author}{Welsch, R.},
  \bibinfo{author}{Allodi, M.~A.}, \bibinfo{author}{Miller, T.~F.} \&
  \bibinfo{author}{Blake, G.~A.}
\newblock \bibinfo{title}{{2D THz-THz-Raman Photon-Echo Spectroscopy} of
  {Molecular Vibrations} in {Liquid Bromoform}}.
\newblock \emph{\bibinfo{journal}{J. Phys. Chem. Lett.}}
  \textbf{\bibinfo{volume}{8}}, \bibinfo{pages}{4640--4644}
  (\bibinfo{year}{2017}).

\bibitem{Magdau2019}
\bibinfo{author}{Magd{\u a}u, I.~B.}, \bibinfo{author}{Mead, G.~J.},
  \bibinfo{author}{Blake, G.~A.} \& \bibinfo{author}{Miller, T.~F.}
\newblock \bibinfo{title}{Interpretation of the {THz-THz-Raman Spectrum} of
  {Bromoform}}.
\newblock \emph{\bibinfo{journal}{J. Phys. Chem. A}}
  \textbf{\bibinfo{volume}{123}}, \bibinfo{pages}{7278--7287}
  (\bibinfo{year}{2019}).

\bibitem{Mead2020}
\bibinfo{author}{Mead, G.}, \bibinfo{author}{Lin, H.~W.},
  \bibinfo{author}{Magd{\u a}u, I.~B.}, \bibinfo{author}{Miller, T.~F.} \&
  \bibinfo{author}{Blake, G.~A.}
\newblock \bibinfo{title}{Sum-{Frequency Signals} in
  {2D-Terahertz-Terahertz-Raman Spectroscopy}}.
\newblock \emph{\bibinfo{journal}{J. Phys. Chem. B}}
  \textbf{\bibinfo{volume}{124}}, \bibinfo{pages}{8904--8908}
  (\bibinfo{year}{2020}).

\bibitem{Ghalgaoui2020Water}
\bibinfo{author}{Ghalgaoui, A.} \emph{et~al.}
\newblock \bibinfo{title}{Field-{Induced Tunneling Ionization} and
  {Terahertz-Driven Electron Dynamics} in {Liquid Water}}.
\newblock \emph{\bibinfo{journal}{J. Phys. Chem. Lett.}}
  \textbf{\bibinfo{volume}{11}}, \bibinfo{pages}{7717--7722}
  (\bibinfo{year}{2020}).

\bibitem{Runge2023Sol}
\bibinfo{author}{Runge, M.}, \bibinfo{author}{Reimann, K.},
  \bibinfo{author}{Woerner, M.} \& \bibinfo{author}{Elsaesser, T.}
\newblock \bibinfo{title}{Nonlinear {Terahertz Polarizability} of {Electrons
  Solvated} in a {Polar Liquid}}.
\newblock \emph{\bibinfo{journal}{Phys. Rev. Lett.}}
  \textbf{\bibinfo{volume}{131}}, \bibinfo{pages}{166902}
  (\bibinfo{year}{2023}).

\bibitem{Johnson2019}
\bibinfo{author}{Johnson, C.~L.}, \bibinfo{author}{Knighton, B.~E.} \&
  \bibinfo{author}{Johnson, J.~A.}
\newblock \bibinfo{title}{Distinguishing {Nonlinear Terahertz Excitation
  Pathways} with {Two-Dimensional Spectroscopy}}.
\newblock \emph{\bibinfo{journal}{Phys. Rev. Lett.}}
  \textbf{\bibinfo{volume}{122}}, \bibinfo{pages}{073901}
  (\bibinfo{year}{2019}).

\bibitem{Biggs2023}
\bibinfo{author}{Biggs, M.~F.} \emph{et~al.}
\newblock \bibinfo{title}{Pump {Pulse Bandwidth-Activated Nonlinear Phononic
  Coupling} in {CdWO}$_4$} (\bibinfo{year}{2023}).
\newblock \eprint{arXiv:2310.08747}.

\bibitem{Runge2023Bis}
\bibinfo{author}{Runge, M.} \emph{et~al.}
\newblock \bibinfo{title}{{Ultrafast Carrier Dynamics and Symmetry Reduction in
  Bismuth by Nonperturbative Optical Excitation in the Terahertz Range}}.
\newblock \emph{\bibinfo{journal}{Phys. Rev. B}}
  \textbf{\bibinfo{volume}{107}}, \bibinfo{pages}{245140}
  (\bibinfo{year}{2023}).

\bibitem{mukamel1995principles}
\bibinfo{author}{Mukamel, S.}
\newblock \emph{\bibinfo{title}{Principles of Nonlinear Optical Spectroscopy}}
  Oxford series in optical and imaging sciences (\bibinfo{publisher}{Oxford
  University Press}, \bibinfo{year}{1995}).

\bibitem{Wan2019}
\bibinfo{author}{Wan, Y.} \& \bibinfo{author}{Armitage, N.~P.}
\newblock \bibinfo{title}{{Resolving Continua of Fractional Excitations by
  Spinon Echo in THz 2D Coherent Spectroscopy}}.
\newblock \emph{\bibinfo{journal}{Phys. Rev. Lett.}}
  \textbf{\bibinfo{volume}{122}}, \bibinfo{pages}{257401}
  (\bibinfo{year}{2019}).

\bibitem{Nandkishore2021}
\bibinfo{author}{Nandkishore, R.~M.}, \bibinfo{author}{Choi, W.} \&
  \bibinfo{author}{Kim, Y.~B.}
\newblock \bibinfo{title}{Spectroscopic fingerprints of gapped quantum spin
  liquids, both conventional and fractonic}.
\newblock \emph{\bibinfo{journal}{Phys. Rev. Res.}}
  \textbf{\bibinfo{volume}{3}}, \bibinfo{pages}{013254} (\bibinfo{year}{2021}).

\bibitem{Choi2020}
\bibinfo{author}{Choi, W.}, \bibinfo{author}{Lee, K.~H.} \&
  \bibinfo{author}{Kim, Y.~B.}
\newblock \bibinfo{title}{Theory of two-dimensional nonlinear spectroscopy for
  the {Kitaev} spin liquid}.
\newblock \emph{\bibinfo{journal}{Phys. Rev. Lett.}}
  \textbf{\bibinfo{volume}{124}}, \bibinfo{pages}{117205}
  (\bibinfo{year}{2020}).

\bibitem{Hu1990}
\bibinfo{author}{Hu, Y.~Z.} \emph{et~al.}
\newblock \bibinfo{title}{Biexcitons in semiconductor quantum dots}.
\newblock \emph{\bibinfo{journal}{Phys. Rev. Lett.}}
  \textbf{\bibinfo{volume}{64}}, \bibinfo{pages}{1805--1807}
  (\bibinfo{year}{1990}).

\bibitem{You2015}
\bibinfo{author}{You, Y.} \emph{et~al.}
\newblock \bibinfo{title}{Observation of biexcitons in monolayer {WSe$_2$}}.
\newblock \emph{\bibinfo{journal}{Nat. Phys.}} \textbf{\bibinfo{volume}{11}},
  \bibinfo{pages}{477--481} (\bibinfo{year}{2015}).

\bibitem{stone2009two}
\bibinfo{author}{Stone, K.~W.} \emph{et~al.}
\newblock \bibinfo{title}{Two-quantum {2D FT} electronic spectroscopy of
  biexcitons in {GaAs} quantum wells}.
\newblock \emph{\bibinfo{journal}{Science}} \textbf{\bibinfo{volume}{324}},
  \bibinfo{pages}{1169--1173} (\bibinfo{year}{2009}).

\bibitem{karaiskaj2010two}
\bibinfo{author}{Karaiskaj, D.} \emph{et~al.}
\newblock \bibinfo{title}{{Two-Quantum Many-Body Coherences in Two-Dimensional
  Fourier-Transform Spectra of Exciton Resonances in Semiconductor Quantum
  Wells}}.
\newblock \emph{\bibinfo{journal}{Phys. Rev. Lett.}}
  \textbf{\bibinfo{volume}{104}}, \bibinfo{pages}{117401}
  (\bibinfo{year}{2010}).

\bibitem{Wortis1963}
\bibinfo{author}{Wortis, M.}
\newblock \bibinfo{title}{{Bound States of Two Spin Waves in the Heisenberg
  Ferromagnet}}.
\newblock \emph{\bibinfo{journal}{Phys. Rev.}} \textbf{\bibinfo{volume}{132}},
  \bibinfo{pages}{85--97} (\bibinfo{year}{1963}).

\bibitem{Lorenzana1995}
\bibinfo{author}{Lorenzana, J.} \& \bibinfo{author}{Sawatzky, G.~A.}
\newblock \bibinfo{title}{Theory of phonon-assisted multimagnon optical
  absorption and bimagnon states in quantum antiferromagnets}.
\newblock \emph{\bibinfo{journal}{Phys. Rev. B}} \textbf{\bibinfo{volume}{52}},
  \bibinfo{pages}{9576--9589} (\bibinfo{year}{1995}).

\bibitem{Cohen1969}
\bibinfo{author}{Cohen, M.~H.} \& \bibinfo{author}{Ruvalds, J.}
\newblock \bibinfo{title}{Two-phonon bound states}.
\newblock \emph{\bibinfo{journal}{Phys. Rev. Lett.}}
  \textbf{\bibinfo{volume}{23}}, \bibinfo{pages}{1378--1381}
  (\bibinfo{year}{1969}).

\bibitem{Shahbazyan2000}
\bibinfo{author}{Shahbazyan, T.}, \bibinfo{author}{Primozich, N.} \&
  \bibinfo{author}{Perakis, I.}
\newblock \bibinfo{title}{Ultrafast {Coulomb-induced} dynamics of quantum well
  magnetoexcitons}.
\newblock \emph{\bibinfo{journal}{Phys. Rev. B}} \textbf{\bibinfo{volume}{62}},
  \bibinfo{pages}{15925} (\bibinfo{year}{2000}).

\bibitem{Klein1980}
\bibinfo{author}{Sooryakumar, R.} \& \bibinfo{author}{Klein, M.~V.}
\newblock \bibinfo{title}{Raman scattering by superconducting-gap excitations
  and their coupling to charge-density waves}.
\newblock \emph{\bibinfo{journal}{Phys. Rev. Lett.}}
  \textbf{\bibinfo{volume}{45}}, \bibinfo{pages}{660--662}
  (\bibinfo{year}{1980}).

\bibitem{Littlewood1981}
\bibinfo{author}{Littlewood, P.~B.} \& \bibinfo{author}{Varma, C.~M.}
\newblock \bibinfo{title}{Gauge-invariant theory of the dynamical interaction
  of charge density waves and superconductivity}.
\newblock \emph{\bibinfo{journal}{Phys. Rev. Lett.}}
  \textbf{\bibinfo{volume}{47}}, \bibinfo{pages}{811--814}
  (\bibinfo{year}{1981}).

\bibitem{Podolsky2011}
\bibinfo{author}{Podolsky, D.}, \bibinfo{author}{Auerbach, A.} \&
  \bibinfo{author}{Arovas, D.~P.}
\newblock \bibinfo{title}{Visibility of the amplitude ({Higgs}) mode in
  condensed matter}.
\newblock \emph{\bibinfo{journal}{Phys. Rev. B}} \textbf{\bibinfo{volume}{84}},
  \bibinfo{pages}{174522} (\bibinfo{year}{2011}).

\bibitem{Pekker2015}
\bibinfo{author}{Pekker, D.} \& \bibinfo{author}{Varma, C.}
\newblock \bibinfo{title}{Amplitude/{Higgs} modes in condensed matter physics}.
\newblock \emph{\bibinfo{journal}{Annu. Rev. Condens. Matter Phys.}}
  \textbf{\bibinfo{volume}{6}}, \bibinfo{pages}{269--297}
  (\bibinfo{year}{2015}).

\bibitem{krull2016}
\bibinfo{author}{Krull, H.}, \bibinfo{author}{Bittner, N.},
  \bibinfo{author}{Uhrig, G.}, \bibinfo{author}{Manske, D.} \&
  \bibinfo{author}{Schnyder, A.}
\newblock \bibinfo{title}{Coupling of {Higgs} and {Leggett} modes in
  non-equilibrium superconductors}.
\newblock \emph{\bibinfo{journal}{Nat. Commun.}} \textbf{\bibinfo{volume}{7}},
  \bibinfo{pages}{11921} (\bibinfo{year}{2016}).

\bibitem{Cea2016}
\bibinfo{author}{Cea, T.}, \bibinfo{author}{Castellani, C.} \&
  \bibinfo{author}{Benfatto, L.}
\newblock \bibinfo{title}{Nonlinear optical effects and third-harmonic
  generation in superconductors: {Cooper} pairs versus {Higgs} mode
  contribution}.
\newblock \emph{\bibinfo{journal}{Phys. Rev. B}} \textbf{\bibinfo{volume}{93}},
  \bibinfo{pages}{180507(R)} (\bibinfo{year}{2016}).

\bibitem{Aoki2017}
\bibinfo{author}{Murotani, Y.}, \bibinfo{author}{Tsuji, N.} \&
  \bibinfo{author}{Aoki, H.}
\newblock \bibinfo{title}{Theory of light-induced resonances with collective
  {Higgs} and {Leggett} modes in multiband superconductors}.
\newblock \emph{\bibinfo{journal}{Phys. Rev. B}} \textbf{\bibinfo{volume}{95}},
  \bibinfo{pages}{104503} (\bibinfo{year}{2017}).

\bibitem{Wu2019}
\bibinfo{author}{Yang, F.} \& \bibinfo{author}{Wu, M.~W.}
\newblock \bibinfo{title}{Gauge-invariant microscopic kinetic theory of
  superconductivity: Application to the optical response of {Nambu-Goldstone}
  and {Higgs} modes}.
\newblock \emph{\bibinfo{journal}{Phys. Rev. B}}
  \textbf{\bibinfo{volume}{100}}, \bibinfo{pages}{104513}
  (\bibinfo{year}{2019}).

\bibitem{Schwarz2020}
\bibinfo{author}{Schwarz, L.}, \bibinfo{author}{Fauseweh, B.} \&
  \bibinfo{author}{Tsuji, N. e.~a.}
\newblock \bibinfo{title}{Classification and characterization of nonequilibrium
  {Higgs} modes in unconventional superconductors.}
\newblock \emph{\bibinfo{journal}{Nat. Commun.}} \textbf{\bibinfo{volume}{11}},
  \bibinfo{pages}{287} (\bibinfo{year}{2020}).

\bibitem{Klein2010}
\bibinfo{author}{Klein, M.~V.}
\newblock \bibinfo{title}{Theory of {Raman} scattering from {Leggett's}
  collective mode in a multiband superconductor: Application to
  {$\text{MgB}_{2}$}}.
\newblock \emph{\bibinfo{journal}{Phys. Rev. B}} \textbf{\bibinfo{volume}{82}},
  \bibinfo{pages}{014507} (\bibinfo{year}{2010}).

\bibitem{Blumberg}
\bibinfo{author}{Blumberg, G.} \emph{et~al.}
\newblock \bibinfo{title}{Observation of {Leggett's} collective mode in a
  multiband {${\mathrm{MgB}}_{2}$} superconductor}.
\newblock \emph{\bibinfo{journal}{Phys. Rev. Lett.}}
  \textbf{\bibinfo{volume}{99}}, \bibinfo{pages}{227002}
  (\bibinfo{year}{2007}).

\bibitem{Giorgianni2019}
\bibinfo{author}{Giorgianni, F.} \emph{et~al.}
\newblock \bibinfo{title}{Leggett mode controlled by light pulses}.
\newblock \emph{\bibinfo{journal}{Nat. Phys.}} \textbf{\bibinfo{volume}{15}},
  \bibinfo{pages}{341--346} (\bibinfo{year}{2019}).

\bibitem{Shimano2019}
\bibinfo{author}{Nakamura, S.} \emph{et~al.}
\newblock \bibinfo{title}{Infrared activation of the {Higgs} mode by
  supercurrent injection in superconducting {NbN}}.
\newblock \emph{\bibinfo{journal}{Phys. Rev. Lett.}}
  \textbf{\bibinfo{volume}{122}}, \bibinfo{pages}{257001}
  (\bibinfo{year}{2019}).

\bibitem{seibold2021}
\bibinfo{author}{Seibold, G.}, \bibinfo{author}{Udina, M.},
  \bibinfo{author}{Castellani, C.} \& \bibinfo{author}{Benfatto, L.}
\newblock \bibinfo{title}{Third harmonic generation from collective modes in
  disordered superconductors}.
\newblock \emph{\bibinfo{journal}{Phys. Rev. B}}
  \textbf{\bibinfo{volume}{103}}, \bibinfo{pages}{014512}
  (\bibinfo{year}{2021}).

\bibitem{udina2022thz}
\bibinfo{author}{Udina, M.} \emph{et~al.}
\newblock \bibinfo{title}{{THz} non-linear optical response in cuprates:
  predominance of the {BCS} response over the {Higgs} mode}.
\newblock \emph{\bibinfo{journal}{Faraday Discuss.}}
  \textbf{\bibinfo{volume}{237}}, \bibinfo{pages}{168--185}
  (\bibinfo{year}{2022}).

\bibitem{yuan2024}
\bibinfo{author}{Yuan, J.} \emph{et~al.}
\newblock \bibinfo{title}{Selective excitation of collective modes in multiband
  superconductor {MgB}$_2$} (\bibinfo{year}{2024}).
\newblock \eprint{arXiv:2412.13830}.

\bibitem{Matsunaga:2012}
\bibinfo{author}{Matsunaga, R.} \& \bibinfo{author}{Shimano, R.}
\newblock \bibinfo{title}{Nonequilibrium {BCS} state dynamics induced by
  intense terahertz pulses in a superconducting {NbN} film}.
\newblock \emph{\bibinfo{journal}{Phys. Rev. Lett.}}
  \textbf{\bibinfo{volume}{109}}, \bibinfo{pages}{187002}
  (\bibinfo{year}{2012}).

\bibitem{Matsunaga:2013}
\bibinfo{author}{Matsunaga, R.} \emph{et~al.}
\newblock \bibinfo{title}{{Higgs} amplitude mode in the {BCS} superconductors
  {$\mathrm{Nb}_{1\mathrm{\text{\ensuremath{-}}}x}\mathrm{Ti}_{x}\mathrm{N}$}
  induced by terahertz pulse excitation}.
\newblock \emph{\bibinfo{journal}{Phys. Rev. Lett.}}
  \textbf{\bibinfo{volume}{111}}, \bibinfo{pages}{057002}
  (\bibinfo{year}{2013}).

\bibitem{matsunaga2014}
\bibinfo{author}{Matsunaga, R.} \emph{et~al.}
\newblock \bibinfo{title}{Light-induced collective pseudospin precession
  resonating with {Higgs} mode in a superconductor}.
\newblock \emph{\bibinfo{journal}{Science}} \textbf{\bibinfo{volume}{345}},
  \bibinfo{pages}{1145--1149} (\bibinfo{year}{2014}).

\bibitem{Matsunaga2017}
\bibinfo{author}{Matsunaga, R.} \emph{et~al.}
\newblock \bibinfo{title}{Polarization-resolved terahertz third-harmonic
  generation in a single-crystal superconductor {NbN}: Dominance of the {Higgs}
  mode beyond the {BCS} approximation}.
\newblock \emph{\bibinfo{journal}{Phys. Rev. B}} \textbf{\bibinfo{volume}{96}},
  \bibinfo{pages}{020505(R)} (\bibinfo{year}{2017}).

\bibitem{Yang2019b}
\bibinfo{author}{Yang, X.} \emph{et~al.}
\newblock \bibinfo{title}{Ultrafast nonthermal terahertz electrodynamics and
  possible quantum energy transfer in the {$\mathrm{Nb}_{3}\mathrm{Sn}$}
  superconductor}.
\newblock \emph{\bibinfo{journal}{Phys. Rev. B}} \textbf{\bibinfo{volume}{99}},
  \bibinfo{pages}{094504} (\bibinfo{year}{2019}).

\bibitem{Chu2020}
\bibinfo{author}{Chu, H.} \emph{et~al.}
\newblock \bibinfo{title}{Phase-resolved {Higgs} response in superconducting
  cuprates}.
\newblock \emph{\bibinfo{journal}{Nat. Commun.}} \textbf{\bibinfo{volume}{11}},
  \bibinfo{pages}{1793} (\bibinfo{year}{2020}).

\bibitem{hybrid-higgs}
\bibinfo{author}{Vaswani, C.} \emph{et~al.}
\newblock \bibinfo{title}{Light quantum control of persisting {Higgs} modes in
  iron-based superconductors}.
\newblock \emph{\bibinfo{journal}{Nat. Commun.}} \textbf{\bibinfo{volume}{12}},
  \bibinfo{pages}{258} (\bibinfo{year}{2021}).

\bibitem{cheng2023lowenergy}
\bibinfo{author}{Cheng, B.} \emph{et~al.}
\newblock \bibinfo{title}{Evidence for $d$-wave superconductivity of
  infinite-layer nickelates from low-energy electrodynamics}.
\newblock \emph{\bibinfo{journal}{Nat. Mater.}} \textbf{\bibinfo{volume}{23}},
  \bibinfo{pages}{775--781} (\bibinfo{year}{2024}).

\bibitem{salvador2025}
\bibinfo{author}{Salvador, A.~G.} \emph{et~al.}
\newblock \bibinfo{title}{Two-dimensional spectroscopy of bosonic collective
  excitations in disordered many-body systems} (\bibinfo{year}{2025}).
\newblock \eprint{arXiv:2501.16856}.

\bibitem{Manske2023}
\bibinfo{author}{Puviani, M.}, \bibinfo{author}{Haenel, R.} \&
  \bibinfo{author}{Manske, D.}
\newblock \bibinfo{title}{Quench-drive spectroscopy and high-harmonic
  generation in {BCS} superconductors}.
\newblock \emph{\bibinfo{journal}{Phys. Rev. B}}
  \textbf{\bibinfo{volume}{107}}, \bibinfo{pages}{094501}
  (\bibinfo{year}{2023}).

\bibitem{Puviani2024}
\bibinfo{author}{Puviani, M.}
\newblock \bibinfo{title}{Theory of symmetry-resolved quench-drive
  spectroscopy: Nonlinear response of phase-fluctuating superconductors}.
\newblock \emph{\bibinfo{journal}{Phys. Rev. B}}
  \textbf{\bibinfo{volume}{109}}, \bibinfo{pages}{214515}
  (\bibinfo{year}{2024}).

\bibitem{Mootz2022}
\bibinfo{author}{Mootz, M.}, \bibinfo{author}{Luo, L.}, \bibinfo{author}{Wang,
  J.} \& \bibinfo{author}{Perakis, l.~E.}
\newblock \bibinfo{title}{Visualization and quantum control of
  light-accelerated condensates by terahertz multi-dimensional coherent
  spectroscopy}.
\newblock \emph{\bibinfo{journal}{Commun. Phys.}} \textbf{\bibinfo{volume}{5}},
  \bibinfo{pages}{47} (\bibinfo{year}{2022}).

\bibitem{Zhou2021}
\bibinfo{author}{Zhou, X.} \emph{et~al.}
\newblock \bibinfo{title}{High-temperature superconductivity}.
\newblock \emph{\bibinfo{journal}{Nat. Rev. Phys.}}
  \textbf{\bibinfo{volume}{3}}, \bibinfo{pages}{462--465}
  (\bibinfo{year}{2021}).

\bibitem{Fausti2011}
\bibinfo{author}{Fausti, D.} \emph{et~al.}
\newblock \bibinfo{title}{Light-induced superconductivity in a stripe-ordered
  cuprate}.
\newblock \emph{\bibinfo{journal}{Science}} \textbf{\bibinfo{volume}{331}},
  \bibinfo{pages}{189--191} (\bibinfo{year}{2011}).

\bibitem{Mitrano2016}
\bibinfo{author}{Mitrano, M.} \emph{et~al.}
\newblock \bibinfo{title}{Possible light-induced superconductivity in
  {K$_3$C$_{60}$} at high temperature}.
\newblock \emph{\bibinfo{journal}{Nature}} \textbf{\bibinfo{volume}{530}},
  \bibinfo{pages}{461--464} (\bibinfo{year}{2016}).

\bibitem{Budden2021}
\bibinfo{author}{Budden, M.} \emph{et~al.}
\newblock \bibinfo{title}{Evidence for metastable photo-induced
  superconductivity in {K$_3$C$_{60}$}}.
\newblock \emph{\bibinfo{journal}{Nat. Phys.}} \textbf{\bibinfo{volume}{17}},
  \bibinfo{pages}{611--618} (\bibinfo{year}{2021}).

\bibitem{Rowe2023}
\bibinfo{author}{Rowe, E.} \emph{et~al.}
\newblock \bibinfo{title}{Resonant enhancement of photo-induced
  superconductivity in {K$_3$C$_{60}$}}.
\newblock \emph{\bibinfo{journal}{Nat. Phys.}} \textbf{\bibinfo{volume}{19}},
  \bibinfo{pages}{1821--1826} (\bibinfo{year}{2023}).

\bibitem{Chattopadhyay2025}
\bibinfo{author}{Chattopadhyay, S.} \emph{et~al.}
\newblock \bibinfo{title}{Metastable photo-induced superconductivity far above
  {$T_\text{c}$}}.
\newblock \emph{\bibinfo{journal}{npj Quantum Mater.}}
  \textbf{\bibinfo{volume}{10}}, \bibinfo{pages}{34} (\bibinfo{year}{2025}).

\bibitem{Kirilyuk2010}
\bibinfo{author}{Kirilyuk, A.}, \bibinfo{author}{Kimel, A.~V.} \&
  \bibinfo{author}{Rasing, T.}
\newblock \bibinfo{title}{Ultrafast optical manipulation of magnetic order}.
\newblock \emph{\bibinfo{journal}{Rev. Mod. Phys.}}
  \textbf{\bibinfo{volume}{82}}, \bibinfo{pages}{2731--2784}
  (\bibinfo{year}{2010}).

\bibitem{Kampfrath2011}
\bibinfo{author}{Kampfrath, T.} \emph{et~al.}
\newblock \bibinfo{title}{Coherent terahertz control of antiferromagnetic spin
  waves}.
\newblock \emph{\bibinfo{journal}{Nat. Photonics}}
  \textbf{\bibinfo{volume}{5}}, \bibinfo{pages}{31--34} (\bibinfo{year}{2011}).

\bibitem{Baierl2016}
\bibinfo{author}{Baierl, S.} \emph{et~al.}
\newblock \bibinfo{title}{Nonlinear spin control by terahertz-driven anisotropy
  fields}.
\newblock \emph{\bibinfo{journal}{Nat. Photonics}}
  \textbf{\bibinfo{volume}{10}}, \bibinfo{pages}{715--718}
  (\bibinfo{year}{2016}).

\bibitem{Afanasiev2021}
\bibinfo{author}{Afanasiev, D.} \emph{et~al.}
\newblock \bibinfo{title}{Ultrafast control of magnetic interactions via
  light-driven phonons}.
\newblock \emph{\bibinfo{journal}{Nat. Mater.}} \textbf{\bibinfo{volume}{20}},
  \bibinfo{pages}{607--611} (\bibinfo{year}{2021}).

\bibitem{Behovits2023}
\bibinfo{author}{Behovits, Y.} \emph{et~al.}
\newblock \bibinfo{title}{Terahertz {N{\'e}el} spin-orbit torques drive
  nonlinear magnon dynamics in antiferromagnetic {Mn$_2$Au}}.
\newblock \emph{\bibinfo{journal}{Nat. Commun.}} \textbf{\bibinfo{volume}{14}},
  \bibinfo{pages}{6038} (\bibinfo{year}{2023}).

\bibitem{Chovan2006}
\bibinfo{author}{Chovan, J.}, \bibinfo{author}{Kavousanaki, E.} \&
  \bibinfo{author}{Perakis, I.}
\newblock \bibinfo{title}{Ultrafast light-induced magnetization dynamics of
  ferromagnetic semiconductors}.
\newblock \emph{\bibinfo{journal}{Phys. Rev. Lett.}}
  \textbf{\bibinfo{volume}{96}}, \bibinfo{pages}{057402}
  (\bibinfo{year}{2006}).

\bibitem{Chovan2008}
\bibinfo{author}{Chovan, J.} \& \bibinfo{author}{Perakis, I.}
\newblock \bibinfo{title}{Femtosecond control of the magnetization in
  ferromagnetic semiconductors}.
\newblock \emph{\bibinfo{journal}{Phys. Rev. B}} \textbf{\bibinfo{volume}{77}},
  \bibinfo{pages}{085321} (\bibinfo{year}{2008}).

\bibitem{Wang2009}
\bibinfo{author}{Wang, J.} \emph{et~al.}
\newblock \bibinfo{title}{Memory effects in photoinduced femtosecond
  magnetization rotation in ferromagnetic {GaMnAs}}.
\newblock \emph{\bibinfo{journal}{Appl. Phys. Lett.}}
  \textbf{\bibinfo{volume}{94}}, \bibinfo{pages}{021101}
  (\bibinfo{year}{2009}).

\bibitem{Kapetanakis2009}
\bibinfo{author}{Kapetanakis, M.}, \bibinfo{author}{Perakis, I.},
  \bibinfo{author}{Wickey, K.}, \bibinfo{author}{Piermarocchi, C.} \&
  \bibinfo{author}{Wang, J.}
\newblock \bibinfo{title}{{Femtosecond Coherent Control of Spins in (Ga,Mn)As
  Ferromagnetic Semiconductors Using Light}}.
\newblock \emph{\bibinfo{journal}{Phys. Rev. Lett.}}
  \textbf{\bibinfo{volume}{103}}, \bibinfo{pages}{047404}
  (\bibinfo{year}{2009}).

\bibitem{Lingos2015}
\bibinfo{author}{Lingos, P.}, \bibinfo{author}{Wang, J.} \&
  \bibinfo{author}{Perakis, I.}
\newblock \bibinfo{title}{Manipulating femtosecond spin-orbit torques with
  laser pulse sequences to control magnetic memory states and ringing}.
\newblock \emph{\bibinfo{journal}{Phys. Rev. B}} \textbf{\bibinfo{volume}{91}},
  \bibinfo{pages}{195203} (\bibinfo{year}{2015}).

\bibitem{Patz2015}
\bibinfo{author}{Patz, A.} \emph{et~al.}
\newblock \bibinfo{title}{Ultrafast probes of nonequilibrium hole spin
  relaxation in the ferromagnetic semiconductor {GaMnAs}}.
\newblock \emph{\bibinfo{journal}{Phys. Rev. B}} \textbf{\bibinfo{volume}{91}},
  \bibinfo{pages}{155108} (\bibinfo{year}{2015}).

\bibitem{mootz2023twodimensional}
\bibinfo{author}{Mootz, M.} \emph{et~al.}
\newblock \bibinfo{title}{Two-dimensional coherent spectrum of high-spin models
  via a quantum computing approach}.
\newblock \emph{\bibinfo{journal}{Quantum Sci. Technol.}}
  \textbf{\bibinfo{volume}{9}}, \bibinfo{pages}{035054} (\bibinfo{year}{2024}).

\bibitem{Mao2018}
\bibinfo{author}{Mao, H.~K.}, \bibinfo{author}{Chen, X.~J.},
  \bibinfo{author}{Ding, Y.}, \bibinfo{author}{Li, B.} \&
  \bibinfo{author}{Wang, L.}
\newblock \bibinfo{title}{Solids, liquids, and gases under high pressure}.
\newblock \emph{\bibinfo{journal}{Rev. Mod. Phys.}}
  \textbf{\bibinfo{volume}{90}}, \bibinfo{pages}{015007}
  (\bibinfo{year}{2018}).

\bibitem{kono}
\bibinfo{author}{Noe, G.~T.} \emph{et~al.}
\newblock \bibinfo{title}{Single-shot terahertz time-domain spectroscopy in
  pulsed high magnetic fields}.
\newblock \emph{\bibinfo{journal}{Opt. Express}} \textbf{\bibinfo{volume}{24}},
  \bibinfo{pages}{30328} (\bibinfo{year}{2016}).

\bibitem{kim2023visualizing}
\bibinfo{author}{Kim, R. H.~J.} \emph{et~al.}
\newblock \bibinfo{title}{Visualizing heterogeneous dipole fields by terahertz
  light coupling in individual nano-junctions}.
\newblock \emph{\bibinfo{journal}{Commun. Phys.}} \textbf{\bibinfo{volume}{6}},
  \bibinfo{pages}{147} (\bibinfo{year}{2023}).

\bibitem{Tokura2017}
\bibinfo{author}{Tokura, Y.}, \bibinfo{author}{Kawasaki, M.} \&
  \bibinfo{author}{Nagaosa, N.}
\newblock \bibinfo{title}{Emergent functions of quantum materials}.
\newblock \emph{\bibinfo{journal}{Nat. Phys.}} \textbf{\bibinfo{volume}{13}},
  \bibinfo{pages}{1056--1068} (\bibinfo{year}{2017}).

\bibitem{Huebener2021}
\bibinfo{author}{H{\"u}bener, H.} \emph{et~al.}
\newblock \bibinfo{title}{Engineering quantum materials with chiral optical
  cavities}.
\newblock \emph{\bibinfo{journal}{Nat. Mater.}} \textbf{\bibinfo{volume}{20}},
  \bibinfo{pages}{438--442} (\bibinfo{year}{2021}).

\bibitem{Yamamoto2015}
\bibinfo{author}{Yamamoto, A.}, \bibinfo{author}{Takeshita, N.},
  \bibinfo{author}{Terakura, C.} \& \bibinfo{author}{Tokura, Y.}
\newblock \bibinfo{title}{High pressure effects revisited for the cuprate
  superconductor family with highest critical temperature}.
\newblock \emph{\bibinfo{journal}{Nat. Commun.}} \textbf{\bibinfo{volume}{6}},
  \bibinfo{pages}{8990} (\bibinfo{year}{2015}).

\bibitem{Sebastian2012}
\bibinfo{author}{Sebastian, S.~E.}, \bibinfo{author}{Harrison, N.} \&
  \bibinfo{author}{Lonzarich, G.~G.}
\newblock \bibinfo{title}{Towards resolution of the {Fermi} surface in
  underdoped {high-$T_c$} superconductors}.
\newblock \emph{\bibinfo{journal}{Rep. Prog. Phys.}}
  \textbf{\bibinfo{volume}{75}}, \bibinfo{pages}{102501}
  (\bibinfo{year}{2012}).

\bibitem{McMillan2002}
\bibinfo{author}{McMillan, P.~F.}
\newblock \bibinfo{title}{New materials from high-pressure experiments}.
\newblock \emph{\bibinfo{journal}{Nat. Mater.}} \textbf{\bibinfo{volume}{1}},
  \bibinfo{pages}{19--25} (\bibinfo{year}{2002}).

\bibitem{Zhou2016}
\bibinfo{author}{Zhou, Y.} \emph{et~al.}
\newblock \bibinfo{title}{Pressure-induced superconductivity in a
  three-dimensional topological material {ZrTe$_5$}}.
\newblock \emph{\bibinfo{journal}{Proc. Natl. Acad. Sci. U.S.A.}}
  \textbf{\bibinfo{volume}{113}}, \bibinfo{pages}{2904--2909}
  (\bibinfo{year}{2016}).

\bibitem{Dias2017}
\bibinfo{author}{Dias, R.~P.} \& \bibinfo{author}{Silvera, I.~F.}
\newblock \bibinfo{title}{Observation of the {Wigner-Huntington} transition to
  metallic hydrogen}.
\newblock \emph{\bibinfo{journal}{Science}} \textbf{\bibinfo{volume}{355}},
  \bibinfo{pages}{715--718} (\bibinfo{year}{2017}).

\bibitem{JW}
\bibinfo{author}{Park, J.~M.} \emph{et~al.}
\newblock \bibinfo{title}{Ultrafast spectroscopic probing of quasiparticle
  dynamics and charge density wave transitions under high pressure and magnetic
  fields in $\mathrm{Pr}_4\mathrm{Ni}_3\mathrm{O}_{10}$}
  (\bibinfo{year}{2025}).

\bibitem{kim2021terahertz}
\bibinfo{author}{Kim, R. H.~J.} \emph{et~al.}
\newblock \bibinfo{title}{Terahertz nano-imaging of electronic strip
  heterogeneity in a {Dirac} semimetal}.
\newblock \emph{\bibinfo{journal}{ACS Photonics}} \textbf{\bibinfo{volume}{8}},
  \bibinfo{pages}{1873--1880} (\bibinfo{year}{2021}).

\bibitem{kim2022terahertz}
\bibinfo{author}{Kim, R. H.~J.} \emph{et~al.}
\newblock \bibinfo{title}{Terahertz nanoimaging of perovskite solar cell
  materials}.
\newblock \emph{\bibinfo{journal}{ACS Photonics}} \textbf{\bibinfo{volume}{9}},
  \bibinfo{pages}{3550--3556} (\bibinfo{year}{2022}).

\bibitem{Zhang2018}
\bibinfo{author}{Zhang, J.} \emph{et~al.}
\newblock \bibinfo{title}{Terahertz nanoimaging of graphene}.
\newblock \emph{\bibinfo{journal}{ACS Photonics}} \textbf{\bibinfo{volume}{5}},
  \bibinfo{pages}{2645--2651} (\bibinfo{year}{2018}).

\bibitem{Guo2024}
\bibinfo{author}{Guo, X.} \emph{et~al.}
\newblock \bibinfo{title}{Terahertz nanoscopy: Advances, challenges, and the
  road ahead}.
\newblock \emph{\bibinfo{journal}{Appl. Phys. Rev.}}
  \textbf{\bibinfo{volume}{11}}, \bibinfo{pages}{021306}
  (\bibinfo{year}{2024}).

\bibitem{kim2023sub}
\bibinfo{author}{Kim, R. H.~J.}, \bibinfo{author}{Park, J.},
  \bibinfo{author}{Haeuser, S.~J.}, \bibinfo{author}{Luo, L.} \&
  \bibinfo{author}{Wang, J.}
\newblock \bibinfo{title}{A sub-2 {Kelvin} cryogenic magneto-terahertz
  scattering-type scanning near-field optical microscope {(cm-THz-sSNOM)}}.
\newblock \emph{\bibinfo{journal}{Rev. Sci. Instrum.}}
  \textbf{\bibinfo{volume}{94}}, \bibinfo{pages}{043702}
  (\bibinfo{year}{2023}).

\bibitem{negahdariNonlinearResponseKitaev2023}
\bibinfo{author}{Negahdari, M.~K.} \& \bibinfo{author}{Langari, A.}
\newblock \bibinfo{title}{Nonlinear response of the {Kitaev} honeycomb lattice
  model in a weak magnetic field}.
\newblock \emph{\bibinfo{journal}{Phys. Rev. B}}
  \textbf{\bibinfo{volume}{107}}, \bibinfo{pages}{134404}
  (\bibinfo{year}{2023}).

\bibitem{qiang2023probing}
\bibinfo{author}{Qiang, Y.}, \bibinfo{author}{Quito, V.~L.},
  \bibinfo{author}{Trevisan, T.~V.} \& \bibinfo{author}{Orth, P.~P.}
\newblock \bibinfo{title}{{Probing Majorana Wave Functions in Kitaev Honeycomb
  Spin Liquids with Second-Order Two-Dimensional Spectroscopy}}.
\newblock \emph{\bibinfo{journal}{Phys. Rev. Lett.}}
  \textbf{\bibinfo{volume}{133}}, \bibinfo{pages}{126505}
  (\bibinfo{year}{2024}).

\bibitem{Potts2024}
\bibinfo{author}{Potts, M.}, \bibinfo{author}{Moessner, R.} \&
  \bibinfo{author}{Benton, O.}
\newblock \bibinfo{title}{{Signatures of Spinon Dynamics and Phase Structure of
  Dipolar-Octupolar Quantum Spin Ices in Two-Dimensional Coherent
  Spectroscopy}}.
\newblock \emph{\bibinfo{journal}{Phys. Rev. Lett.}}
  \textbf{\bibinfo{volume}{133}}, \bibinfo{pages}{226701}
  (\bibinfo{year}{2024}).

\bibitem{zhang2024disentanglingspinexcitationcontinua}
\bibinfo{author}{Zhang, E.~Z.}, \bibinfo{author}{Hickey, C.} \&
  \bibinfo{author}{Kim, Y.~B.}
\newblock \bibinfo{title}{Disentangling spin excitation continua in classical
  and quantum magnets using two-dimensional nonlinear spectroscopy}.
\newblock \emph{\bibinfo{journal}{Phys. Rev. B}}
  \textbf{\bibinfo{volume}{110}}, \bibinfo{pages}{104415}
  (\bibinfo{year}{2024}).

\bibitem{liPhotonEchoLensing2021}
\bibinfo{author}{Li, Z.~L.}, \bibinfo{author}{Oshikawa, M.} \&
  \bibinfo{author}{Wan, Y.}
\newblock \bibinfo{title}{Photon {Echo} from {Lensing} of {Fractional
  Excitations} in {Tomonaga-Luttinger Spin Liquid}}.
\newblock \emph{\bibinfo{journal}{Phys. Rev. X}} \textbf{\bibinfo{volume}{11}},
  \bibinfo{pages}{031035} (\bibinfo{year}{2021}).

\bibitem{hartExtractingSpinonSelfenergies2023}
\bibinfo{author}{Hart, O.} \& \bibinfo{author}{Nandkishore, R.}
\newblock \bibinfo{title}{Extracting spinon self-energies from two-dimensional
  coherent spectroscopy}.
\newblock \emph{\bibinfo{journal}{Phys. Rev. B}}
  \textbf{\bibinfo{volume}{107}}, \bibinfo{pages}{205143}
  (\bibinfo{year}{2023}).

\bibitem{gaoTwodimensionalCoherentSpectrum2023}
\bibinfo{author}{Gao, Q.}, \bibinfo{author}{Liu, Y.}, \bibinfo{author}{Liao,
  H.} \& \bibinfo{author}{Wan, Y.}
\newblock \bibinfo{title}{Two-dimensional coherent spectrum of interacting
  spinons from matrix product states}.
\newblock \emph{\bibinfo{journal}{Phys. Rev. B}}
  \textbf{\bibinfo{volume}{107}}, \bibinfo{pages}{165121}
  (\bibinfo{year}{2023}).

\bibitem{simMicroscopicDetailsTwodimensional2023}
\bibinfo{author}{Sim, G.}, \bibinfo{author}{Pollmann, F.} \&
  \bibinfo{author}{Knolle, J.}
\newblock \bibinfo{title}{Microscopic details of two-dimensional spectroscopy
  of one-dimensional quantum {Ising} magnets}.
\newblock \emph{\bibinfo{journal}{Phys. Rev. B}}
  \textbf{\bibinfo{volume}{108}}, \bibinfo{pages}{134423}
  (\bibinfo{year}{2023}).

\bibitem{liPhotonEchoFractional2023}
\bibinfo{author}{Li, Z.~L.} \& \bibinfo{author}{Wan, Y.}
\newblock \bibinfo{title}{Photon echo and fractional excitation lensing of the
  {$S=\frac{1}{2}$} {XY} spin chain}.
\newblock \emph{\bibinfo{journal}{Phys. Rev. B}}
  \textbf{\bibinfo{volume}{108}}, \bibinfo{pages}{165151}
  (\bibinfo{year}{2023}).

\bibitem{pottsExploitingPolarizationDependence2023}
\bibinfo{author}{Potts, M.}, \bibinfo{author}{Moessner, R.} \&
  \bibinfo{author}{Benton, O.}
\newblock \bibinfo{title}{Exploiting polarization dependence in two-dimensional
  coherent spectroscopy: Examples of
  {${\mathrm{Ce}}_{2}{\mathrm{Zr}}_{2}{\mathrm{O}}_{7}$ and
  ${\mathrm{Nd}}_{2}{\mathrm{Zr}}_{2}{\mathrm{O}}_{7}$}}.
\newblock \emph{\bibinfo{journal}{Phys. Rev. B}}
  \textbf{\bibinfo{volume}{109}}, \bibinfo{pages}{104435}
  (\bibinfo{year}{2024}).

\bibitem{Watanabe2025}
\bibinfo{author}{Watanabe, Y.}, \bibinfo{author}{Trebst, S.} \&
  \bibinfo{author}{Hickey, C.}
\newblock \bibinfo{title}{Revealing quadrupolar excitations with nonlinear
  spectroscopy}.
\newblock \emph{\bibinfo{journal}{Phys. Rev. Lett.}}
  \textbf{\bibinfo{volume}{134}}, \bibinfo{pages}{106703}
  (\bibinfo{year}{2025}).

\bibitem{brenig2025twodimensionalnonlinearopticalresponse}
\bibinfo{author}{Brenig, W.}
\newblock \bibinfo{title}{Two-dimensional nonlinear optical response of a
  spiral magnet} (\bibinfo{year}{2025}).
\newblock \eprint{arXiv:2504.07177}.

\bibitem{Zhang2023DM}
\bibinfo{author}{Zhang, J.} \& \bibinfo{author}{Tanimura, Y.}
\newblock \bibinfo{title}{Coherent two-dimensional {THz} magnetic resonance
  spectroscopies for molecular magnets: Analysis of {Dzyaloshinskii–Moriya}
  interaction}.
\newblock \emph{\bibinfo{journal}{J. Chem. Phys.}}
  \textbf{\bibinfo{volume}{159}}, \bibinfo{pages}{014102}
  (\bibinfo{year}{2023}).

\bibitem{srivastava2025theorynonlinearspectroscopyquantum}
\bibinfo{author}{Srivastava, A.}, \bibinfo{author}{Birnkammer, S.},
  \bibinfo{author}{Sim, G.}, \bibinfo{author}{Knap, M.} \&
  \bibinfo{author}{Knolle, J.}
\newblock \bibinfo{title}{Theory of nonlinear spectroscopy of quantum magnets}
  (\bibinfo{year}{2025}).
\newblock \eprint{arXiv:2502.17554}.

\bibitem{parameswaranAsymptoticallyExactTheory2020}
\bibinfo{author}{Parameswaran, S.~A.} \& \bibinfo{author}{Gopalakrishnan, S.}
\newblock \bibinfo{title}{Asymptotically {Exact Theory} for {Nonlinear
  Spectroscopy} of {Random Quantum Magnets}}.
\newblock \emph{\bibinfo{journal}{Phys. Rev. Lett.}}
  \textbf{\bibinfo{volume}{125}}, \bibinfo{pages}{237601}
  (\bibinfo{year}{2020}).

\bibitem{Takahashi2012}
\bibinfo{author}{Takahashi, Y.}, \bibinfo{author}{Shimano, R.},
  \bibinfo{author}{Kaneko, Y.}, \bibinfo{author}{Murakawa, H.} \&
  \bibinfo{author}{Tokura, Y.}
\newblock \bibinfo{title}{{Magnetoelectric Resonance with Electromagnons in a
  Perovskite Helimagnet}}.
\newblock \emph{\bibinfo{journal}{Nat. Phys.}} \textbf{\bibinfo{volume}{8}},
  \bibinfo{pages}{121--125} (\bibinfo{year}{2012}).

\bibitem{Fiebig2016}
\bibinfo{author}{Fiebig, M.}, \bibinfo{author}{Lottermoser, T.},
  \bibinfo{author}{Meier, D.} \& \bibinfo{author}{Trassin, M.}
\newblock \bibinfo{title}{The evolution of multiferroics}.
\newblock \emph{\bibinfo{journal}{Nat. Rev. Mater.}}
  \textbf{\bibinfo{volume}{1}}, \bibinfo{pages}{16046} (\bibinfo{year}{2016}).

\bibitem{Li2019}
\bibinfo{author}{Li, X.} \emph{et~al.}
\newblock \bibinfo{title}{Terahertz field-induced ferroelectricity in quantum
  paraelectric {SrTiO}{\textsubscript{3}}}.
\newblock \emph{\bibinfo{journal}{Science}} \textbf{\bibinfo{volume}{364}},
  \bibinfo{pages}{1079--1082} (\bibinfo{year}{2019}).

\bibitem{Gao2024}
\bibinfo{author}{Gao, F.~Y.} \emph{et~al.}
\newblock \bibinfo{title}{Giant chiral magnetoelectric oscillations in a van
  der {Waals} multiferroic}.
\newblock \emph{\bibinfo{journal}{Nature}} \textbf{\bibinfo{volume}{632}},
  \bibinfo{pages}{273--279} (\bibinfo{year}{2024}).

\bibitem{Huang2018}
\bibinfo{author}{Huang, B.} \emph{et~al.}
\newblock \bibinfo{title}{Electrical control of {2D} magnetism in bilayer
  {CrI$_3$}}.
\newblock \emph{\bibinfo{journal}{Nat. Nanotechnol.}}
  \textbf{\bibinfo{volume}{13}}, \bibinfo{pages}{544--548}
  (\bibinfo{year}{2018}).

\bibitem{Burch2018}
\bibinfo{author}{Burch, K.~S.}, \bibinfo{author}{Mandrus, D.} \&
  \bibinfo{author}{Park, J.~G.}
\newblock \bibinfo{title}{Magnetism in two-dimensional van der {Waals}
  materials}.
\newblock \emph{\bibinfo{journal}{Nature}} \textbf{\bibinfo{volume}{563}},
  \bibinfo{pages}{47--52} (\bibinfo{year}{2018}).

\bibitem{Gibertini2019}
\bibinfo{author}{Gibertini, M.}, \bibinfo{author}{Koperski, M.},
  \bibinfo{author}{Morpurgo, A.~F.} \& \bibinfo{author}{Novoselov, K.~S.}
\newblock \bibinfo{title}{Magnetic {2D} materials and heterostructures}.
\newblock \emph{\bibinfo{journal}{Nat. Nanotechnol.}}
  \textbf{\bibinfo{volume}{14}}, \bibinfo{pages}{408--419}
  (\bibinfo{year}{2019}).

\bibitem{Gong2019}
\bibinfo{author}{Gong, C.} \& \bibinfo{author}{Zhang, X.}
\newblock \bibinfo{title}{Two-dimensional magnetic crystals and emergent
  heterostructure devices}.
\newblock \emph{\bibinfo{journal}{Science}} \textbf{\bibinfo{volume}{363}},
  \bibinfo{pages}{eaav4450} (\bibinfo{year}{2019}).

\bibitem{femtomag}
\bibinfo{author}{Li, T.} \emph{et~al.}
\newblock \bibinfo{title}{Femtosecond switching of magnetism via strongly
  correlated spin--charge quantum excitations}.
\newblock \emph{\bibinfo{journal}{Nature}} \textbf{\bibinfo{volume}{496}},
  \bibinfo{pages}{69--73} (\bibinfo{year}{2013}).

\bibitem{Jeon2014}
\bibinfo{author}{Jeon, S.} \emph{et~al.}
\newblock \bibinfo{title}{Landau quantization and quasiparticle interference in
  the three-dimensional {Dirac} semimetal {Cd$_3$As$_2$}}.
\newblock \emph{\bibinfo{journal}{Nat. Mater.}} \textbf{\bibinfo{volume}{13}},
  \bibinfo{pages}{851--856} (\bibinfo{year}{2014}).

\bibitem{Otto2017}
\bibinfo{author}{K{\"o}nig-Otto, J.~C.} \emph{et~al.}
\newblock \bibinfo{title}{{Four-Wave Mixing in Landau-Quantized Graphene}}.
\newblock \emph{\bibinfo{journal}{Nano Lett.}} \textbf{\bibinfo{volume}{17}},
  \bibinfo{pages}{2184--2188} (\bibinfo{year}{2017}).

\bibitem{Nenno2020}
\bibinfo{author}{Nenno, D.~M.}, \bibinfo{author}{Garcia, C. A.~C.},
  \bibinfo{author}{Gooth, J.}, \bibinfo{author}{Felser, C.} \&
  \bibinfo{author}{Narang, P.}
\newblock \bibinfo{title}{Axion physics in condensed-matter systems}.
\newblock \emph{\bibinfo{journal}{Nat. Rev. Phys.}}
  \textbf{\bibinfo{volume}{2}}, \bibinfo{pages}{682--696}
  (\bibinfo{year}{2020}).

\bibitem{Qiu2025}
\bibinfo{author}{Qiu, J.~X.} \emph{et~al.}
\newblock \bibinfo{title}{Observation of the axion quasiparticle in {2D}
  {MnBi$_2$Te$_4$}}.
\newblock \emph{\bibinfo{journal}{Nature}} \textbf{\bibinfo{volume}{641}},
  \bibinfo{pages}{62–69} (\bibinfo{year}{2025}).

\bibitem{Weyl_CDW}
\bibinfo{author}{Cheng, B.} \emph{et~al.}
\newblock \bibinfo{title}{Chirality manipulation of ultrafast phase switches in
  a correlated {CDW-Weyl} semimetal}.
\newblock \emph{\bibinfo{journal}{Nat. Commun.}} \textbf{\bibinfo{volume}{15}},
  \bibinfo{pages}{785} (\bibinfo{year}{2024}).

\end{thebibliography}
\end{document}